# The case and context for atmospheric methane as an exoplanet biosignature

Maggie A. Thompson[a,1], Joshua Krissansen-Totton[a], Nicholas Wogan[b], Myriam Telus[c], and Jonathan J. Fortney[a]

Edited by Neta Bahcall, Princeton University, Princeton, NJ; received October 8, 2021; accepted January 31, 2022

**Methane has been proposed as an exoplanet biosignature. Imminent observations with the James Webb Space Telescope may enable methane detections on potentially habitable exoplanets, so it is essential to assess in what planetary contexts methane is a compelling biosignature. Methane's short photochemical lifetime in terrestrial planet atmospheres implies that abundant methane requires large replenishment fluxes. While methane can be produced by a variety of abiotic mechanisms such as outgassing, serpentinizing reactions, and impacts, we argue that—in contrast to an Earth-like biosphere—known abiotic processes cannot easily generate atmospheres rich in $CH_4$ and $CO_2$ with limited CO due to the strong redox disequilibrium between $CH_4$ and $CO_2$. Methane is thus more likely to be biogenic for planets with 1) a terrestrial bulk density, high mean-molecular-weight and anoxic atmosphere, and an old host star; 2) an abundance of $CH_4$ that implies surface fluxes exceeding what could be supplied by abiotic processes; and 3) atmospheric $CO_2$ with comparatively little CO.**

methane | biosignatures | planetary atmospheres

The next phase of exoplanet science will focus on characterizing exoplanet atmospheres, including those of potentially habitable planets. For example, the James Webb Space Telescope (JWST) will be capable of characterizing the atmospheres of transiting, terrestrial planets around low-mass stars, such as the TRAPPIST-1 system (1, 2). A new class of ground-based telescopes (3) may be able to detect atmospheric constituents such as oxygen, water, and carbon dioxide on nearby rocky exoplanets via high-resolution spectroscopy (4). In subsequent decades, the Astro2020 Decadal Survey report has prioritized a large infrared/optical/ultraviolet (UV) telescope built to search for signs of life—biosignatures—on ∼25 habitable-zone planets (5). Life may modify its planetary environment in multiple ways, including producing waste gases that alter a planet's atmospheric composition. As a result, an understanding of detectable biogenic waste gases and their nonbiological false positives is needed.

Terrestrial planets, which are the focus of this study, require significant methane surface fluxes to sustain high atmospheric abundances. On Earth, life sustains large methane surface fluxes, and so methane has long been regarded as a potential biosignature gas for terrestrial exoplanets. Previous studies have considered abiotic methane production (6–11), methane biosignatures in the context of chemical disequilibrium (12–15), and prospects for remote detection of methane in terrestrial atmospheres (6, 9, 15–17). During the Archean eon (4 to 2.5 Ga), Earth's atmosphere likely had high methane abundances (∼$10^2$ to $10^4$ times modern) due to life (i.e., methanogenesis) (8, 18, 19). Methane is thus not a hypothetical biosignature because we know of an inhabited terrestrial planet with detectable levels of biogenic methane—the Archean Earth. However, methane is sometimes dismissed as irredeemably ambiguous due to its ubiquity in planetary environments and potential for nonbiological production (8, 9). Additional work is clearly needed to understand methane biosignatures and their false positives within different planetary contexts.

While other studies have reviewed the biosignature gases oxygen (20), phosphine (21), isoprene (22), and ammonia (23), in the near term, these gases will likely be difficult to detect or will be detectable only in extended $H_2$-dominated atmospheres on planets with large biogenic fluxes. In contrast, for Earth-like biogenic fluxes, methane is one of the few biosignatures that may be readily detectable with JWST (24–26). For example, biological methane on an early Earth-like TRAPPIST-1e could be detectable with 5 to 10 transits with JWST (17, 27) and would remain detectable even with an optically thick aerosol layer at 10 to 100 mbar, assuming plausible instrument noise and negligible stellar contamination (17).

Given the imminent feasibility of observing methane with JWST, it is imperative to determine the planetary conditions where methane is a compelling biosignature. Despite the patchwork of past studies on methane biosignatures, a recent and dedicated investigation

## Significance

Astronomers will soon begin searching for biosignatures, atmospheric gases or surface features produced by life, on potentially habitable planets. Since methane is the only biosignature that the James Webb Space Telescope could readily detect in terrestrial atmospheres, it is imperative to understand methane biosignatures to contextualize these upcoming observations. We explore the necessary planetary context for methane to be a persuasive biosignature and assess whether, and in what planetary environments, abiotic sources of methane could result in false-positive scenarios. With these results, we provide a tentative framework for assessing methane biosignatures. If life is abundant in the universe, then with the correct planetary context, atmospheric methane may be the first detectable indication of life beyond Earth.

Author affiliations: [a]Department of Astronomy and Astrophysics, University of California, Santa Cruz, CA 95064; [b]Department of Earth and Space Sciences, University of Washington, Seattle, WA 98195; and [c]Department of Earth and Planetary Sciences, University of California, Santa Cruz, CA 95064 into the conditions needed for atmospheric methane to be a good exoplanet biosignature is lacking. This study provides an updated assessment of the necessary planetary context for methane biosignatures. First, we present the case for methane as a biosignature, including its short photochemical lifetime and relationship with chemical disequilibrium and CO antibiosignatures. We then explore the possibility of abiotic methane fluxes as large as those caused by known biogenic sources, in part using different modeling tools. We also discuss the purported presence of methane on Mars and simulate atmospheric methane on temperate Titan-like exoplanets. Based on these results, we propose a framework for identifying methane biosignatures and discuss detectability prospects with next-generation missions.

[1]To whom correspondence may be addressed. Email: maapthom@ucsc.edu.

## Biological Methane Production on Earth

The vast majority of methane in Earth's atmosphere today, and throughout most of its history, is biogenic. At present, Earth's ∼30 Tmol/y global methane emissions are predominantly produced directly by life (including anthropogenic sources), and most of the rest is thermogenic methane that derives from previous life, such as metamorphic reactions of organic matter (28). Genuinely abiotic methane emissions, while uncertain, are comparatively tiny (28).

Biological methane production, or methanogenesis, is a simple metabolism performed by anaerobic microbes (i.e., those not requiring oxygen for growth). Methanogenic microbes can be divided into three groups: hydrogenotrophic (reaction **1**), acetoclastic (reaction **2**), and methylotrophic methanogens:

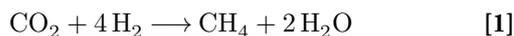
$$CO_2 + 4H_2 \longrightarrow CH_4 + 2H_2O \quad [1]$$

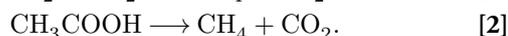
$$CH_3COOH \longrightarrow CH_4 + CO_2. \quad [2]$$

Hydrogenotrophic methanogens typically oxidize $H_2$ and reduce $CO_2$ to $CH_4$ and contribute approximately one-third of current biogenic methane emissions. Acetoclastic methanogens use acetate, contributing approximately two-thirds of current biogenic methane emissions; and finally, methylotrophic methanogens use methylated compounds but do not contribute significantly to global biogenic methane emissions (29). Methane can also be produced indirectly by life as a byproduct of degrading organic matter from dead organisms, called "thermogenic methane."

If life elsewhere is common, methanogenesis may be widespread due to the likely ubiquity of the $CO_2 + H_2$ redox couple in terrestrial planet atmospheres and the potential metabolic payoff from exploiting such commonly outgassed substrates. Methanogenesis is an ancient metabolism on Earth with phylogenetic analyses implying that methanogenesis originated between 4.11 and 3.78 Ga and reconstructions of the last universal common ancestor suggesting methanogens were one of the earliest lifeforms to evolve on Earth (30–32).

There are several reasons to expect methane-cycling biospheres to produce large $CH_4$ fluxes. During the Archean, xenon isotopes—which ostensibly reflect abundances of escaping, hydrogen-bearing species in the upper atmosphere—likely imply large methane abundances (>0.5%) (19, 33). This Xe isotope fractionation can potentially be explained by another hydrogen-bearing species (e.g., >1% $H_2$ or >1% $H_2O$), but such explanations are tentatively disfavored: Catling and Zahnle (19) and Kadoya and Catling (34) place an upper limit of $H_2$ in the Archean atmosphere of 1% and other paleo-pressure and surface temperature estimates likely preclude >1% $H_2O$ above the tropopause. Moreover, multiple ecosystem models for the Archean Earth estimate large biogenic $CH_4$ fluxes and abundant atmospheric $CH_4$ (35–38). Motivated by observations of inefficient methane generation in a ferruginous, sulfate-poor lake ostensibly representative of Precambrian conditions, biogeochemical models of low Precambrian methane have been proposed (39). However, ref. 40 found that such model behavior is dictated by arbitrary forcings and is not compatible with the rock record. In any case, hydrogenotrophic methanogenesis in the Archean water column could maintain substantial $CH_4$ fluxes regardless of organic burial efficiency in sediments (35, 38, 39).

## Results

**The Case for Methane as a Biosignature.** Methane has been highlighted as a potential biosignature gas because it has a short photochemical lifetime (less than ∼1 My) on habitable-zone, rocky planets orbiting solar-type stars. A short photochemical lifetime requires substantial replenishment fluxes to sustain large atmospheric abundances. Methane is removed from an atmosphere photochemically in two ways, depending on the concentration of $CO_2$ relative to $CH_4$ and the presence of other oxidants (41). In the case where $CO_2$ is significantly more abundant, $CH_4$ is destroyed by oxidants and is converted to $CO_2$ (see *SI Appendix*, section 3 for additional reactions):

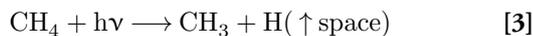
$$CH_4 + h\nu \longrightarrow CH_3 + H(\uparrow space) \quad [3]$$

or

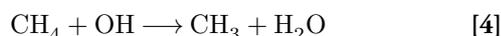
$$CH_4 + OH \longrightarrow CH_3 + H_2O \quad [4]$$

and, subsequently,

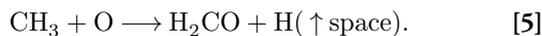
$$CH_3 + O \longrightarrow H_2CO + H(\uparrow space). \quad [5]$$

The C in $H_2CO$ is further oxidized to $CO_2$. The H produced can then be lost to space, thereby irreversibly destroying $CH_4$. Note that OH and O are byproducts of $H_2O$ and $CO_2$ photolysis; an $O_2$-rich atmosphere is not required for rapid $CH_4$ destruction, although it does decrease the $CH_4$ lifetime.

For the case where $CH_4$ is more abundant than $CO_2$, $CH_4$ polymerizes to aerosols, which fall to the ground and remove the atmospheric $CH_4$ (see *SI Appendix*, section 3 for sequence of reactions). If temperatures are high enough in the lower atmosphere, these aerosols could break down and release $CH_4$ back into the atmosphere. In addition, surface deposition and subsequent thermal decomposition in the subsurface could release methane back into the atmosphere. However, some portion of the hydrogen produced by methane photolysis is lost to space, and so, without $H_2$ replenishment, the C:H ratio of condensate material will rise such that the methane is irreversibly lost.

The short atmospheric lifetime of terrestrial planet methane can be quantified. Using the photochemical model PhotochemPy adapted from the Atmos code (42) and created by N. Wogan (43) (*SI Appendix*, section 6A), we explore the stability of atmospheric $CH_4$ for an Archean Earth-like planet (i.e., $N_2$-$CO_2$-$CH_4$) orbiting a 2.7-Ga Sun-like star. Every calculation conserves redox. Consistent with previous studies (7, 13, 44, 45), we find that for atmospheric $CH_4$ mixing ratios greater than ∼$10^{-3}$ to be stable against photochemistry requires replenishing $CH_4$ surface fluxes that are larger than Earth's current biological flux (*SI Appendix*, Fig. S1). If a planet is orbiting a different stellar-type host star, it will be necessary to recalculate the threshold for biological methane surface fluxes. For example, planets orbiting M-stars tend to have lower near-UV radiation compared to Sun-like stars, which reduces the OH produced by $H_2O$ photolysis, permitting higher atmospheric $CH_4$ concentrations (46). Ultimately, however, a terrestrial planet atmosphere that is rich in $CH_4$ cannot persist unless there is a significant replenishment source flux, making it an intriguing candidate for further investigation.

**Methane biosignatures and chemical disequilibrium.** The methane biosignature case is strengthened if its presence in the atmosphere is accompanied by that of a strongly oxidizing companion gas such as $CO_2$ or $O_2/O_3$. This is because it is difficult to explain abundant methane if a terrestrial planet's atmospheric redox state is sufficiently oxidized such that the thermodynamically stable form of carbon is not $CH_4$. Methane in $O_2$-rich atmospheres requires large replenishment fluxes because $CH_4$ and $O_2$ are kinetically unstable and out of thermodynamic equilibrium (47, 48). The kinetic lifetime of methane in $O_2$-rich atmospheres is $\sim 10$ y (44) due to the following net reaction, which is the end result of reactions **3** to **5** above after the $H_2CO$ has been further oxidized to $CO_2$:

$$CH_4 + O_2 \longrightarrow CO_2 + H_2O. \quad [6]$$

Another important thermodynamic disequilibrium is that between $CH_4$ and $CO_2$, which was present on the Archean Earth prior to the rise of $O_2$. Specifically, $CH_4$, $CO_2$, $N_2$, and liquid $H_2O$ coexisted out of equilibrium on the early Earth due to the replenishment of $CH_4$ by life (14). In a weakly reduced Archean atmosphere, $CH_4$'s lifetime would have been short (up to $\sim 2,000$ to $20,000$ y) compared to geologic timescales (49, 50). This short kinetic lifetime of methane does not depend on this thermodynamic disequilibrium with $CO_2$; methane has a short photochemical lifetime in high mean-molecular-weight atmospheres regardless of whether or not $CO_2$ is present in abundance. However, the thermodynamic disequilibrium is of fundamental importance for the discussion of abiotic methane that follows. Crucially, $CH_4$ and $CO_2$ are at opposite ends of the redox spectrum for carbon, separated by eight electrons. This has implications for how both species can be produced via abiotic planetary interior processes, which we explore subsequently; see the discussion of CO below. On the basis of both this thermodynamic disequilibrium and methane's short photochemical lifetime, Krissansen-Totton et al. (14) argued that detecting both abundant $CH_4$ and $CO_2$ in a habitable-zone rocky exoplanet may be a biosignature and, if $CH_4$'s mixing ratio is greater than $\sim 0.001$, the methane is probably biogenic because it is challenging for abiotic sources to sustain large methane fluxes in anoxic atmospheres, similar to the findings of ref. 6.

**CO antibiosignatures and their relationship to CH4 biosignatures.** In the above scenario, the absence of significant atmospheric CO may strengthen the case for biogenic $CH_4$ since 1) microbial life readily consumes CO, a source of free energy, and 2) many abiotic processes that produce $CH_4$ also result in abundant CO (14, 51) (and see below on magmatic outgassing). Life on Earth metabolizes CO because oxidizing it with water yields free energy and because CO metabolism serves as a starting point for carbon fixation (52, 53). Multiple lines of evidence suggest that CO consumption could be a ubiquitous metabolic strategy given its ancient origin on Earth (32, 53–55) and because the required enzymes possess a variety of simple Ni-Fe, Mo, or Cu active sites, suggesting that they have evolved independently multiple times (53, 56, 57). However, the mere presence or absence of CO may not be an unambiguous discriminator between a $CH_4$-producing biosphere and an uninhabited world. An inhabited planet may have $CH_4$, $CO_2$, and some CO in its atmosphere if life is unable to efficiently consume all of the CO (11, 37, 38). In this case, however, the $CO/CH_4$ atmospheric ratio in terrestrial planets' high mean-molecular-weight atmospheres could potentially be used as a diagnostic tool to distinguish anoxic, inhabited planets from lifeless worlds because the $CO/CH_4$ atmospheric ratio reflects the fractional atmospheric free energy that has been exploited.

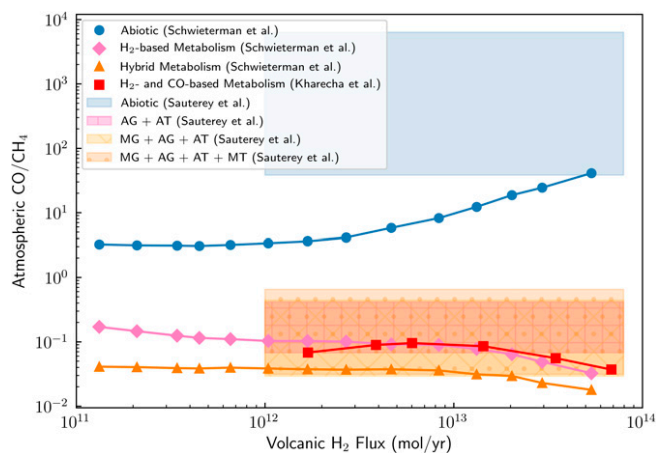

**Fig. 1.** Atmospheric CO to $CH_4$ ratio may help distinguish biogenic and abiotic methane. Shown is ratio of atmospheric CO to $CH_4$ for abiotic worlds and those with biospheres as a function of volcanic $H_2$ flux. The curves show the calculated atmospheric $CO/CH_4$ as a function of volcanic $H_2$ flux for abiotic worlds (blue circles), $H_2$-based biospheres (includes $H_2$-consuming anoxygenic photosynthesis, CO-consuming acetogenesis, organic matter fermentation, and acetotrophic methanogenesis) (pink diamonds), $H_2$-based and Fe-based photosynthesis biospheres (i.e., "hybrid," orange triangles) from ref. 37, and the methanogen–acetogen ecosystem and anoxygenic phototroph–acetogen ecosystem from ref. 35 (i.e., their cases 2 and 3) (red squares). The horizontal shaded regions correspond to the distributions of atmospheric $CO/CH_4$ for abiotic worlds (blue) and those with methanogenic biospheres (pink, yellow, and orange) as a function of volcanic $H_2$ flux calculated by ref. 38. The atmospheric $CO/CH_4$ for abiotic worlds is predicted to be several orders of magnitude greater than that for inhabited worlds. Refs. 35, 37, and 38 found that low $CO/CH_4$ atmospheric ratios ($\sim 0.1$) are a strong sign of methane-cycling biospheres for reducing planets orbiting Sun-like stars like Archean Earth, suggesting that atmospheric $CO/CH_4$ is a good observable diagnostic tool to distinguish abiotic planets from those with anoxic biospheres. The light pink "+"-hatched region corresponds to an ecosystem with CO-based autotrophic acetogens (AG) and methanogenic acetotrophs (AT); the light orange "X"-hatched region corresponds to an ecosystem with $H_2$-based methanogens (MG), AG, and AT; the orange "."-hatched region corresponds to the most complex ecosystem consisting of MG, AG, AT, and anaerobic methanotrophy (MT) (38). All calculations assume a $CO_2$-$CH_4$-$N_2$ bulk atmosphere.

Kharecha et al. (35), Schwieterman et al. (37), and Sauterey et al. (38) found that the atmospheric $CO/CH_4$ ratio for abiotic worlds is predicted to be approximately two orders of magnitude larger than that for inhabited worlds that have anoxic biospheres over a wide range of volcanic $H_2$ fluxes (Fig. 1). Note that we consider only the ecosystems from refs. 35 and 38 where both methanogenesis and CO consumption (acetogenesis plus acetotrophy) have evolved; if these conditions are not met, then larger $CO/CH_4$ ratios are possible, but note the arguments for rapid emergence of CO consumption outlined above. While the atmospheric $CO/CH_4$ ratio is likely an observable parameter that can be used to distinguish lifeless from inhabited, anoxic worlds, additional modeling is required to explore the possible range of $CH_4$, $CO_2$, and CO abundances for a wide variety of biospheres and uninhabited worlds around different host star types.

**Abiotic Sources of Methane.** While the vast majority of Earth's atmospheric methane is produced biotically (28), there are various small abiotic sources of methane that could potentially be enhanced on other planets. Understanding plausible abiotic methane fluxes is necessary for discriminating methane biosignature false-positive scenarios from true signs of metabolism. These abiotic sources can be broadly divided into the following categories (Fig. 2): 1) volcanism and high-temperature magmatic processes,

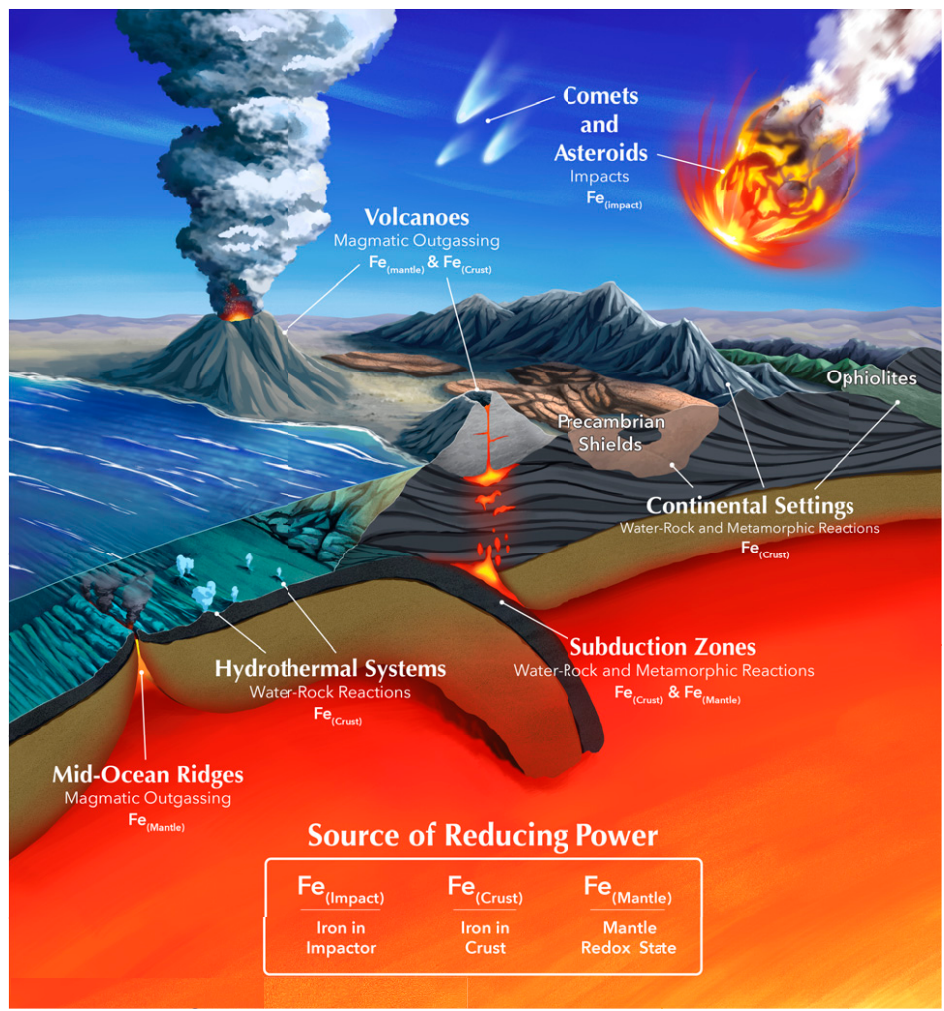

**Fig. 2.** Summary of known abiotic sources of methane on Earth (© 2022 Elena Hartley) (http://www.elabarts.com). In general, the abiotic sources of methane can be divided into three categories: high-temperature magmatic outgassing (volcanism), low-temperature water–rock and metamorphic reactions, and impacts. Currently, subaerial (submarine) volcanoes on Earth generate only $\leq 10^{-3}$ ($\sim 10^{-2}$) Tmol/y of methane (see main text). Low-temperature water–rock reactions that generate methane occur at midocean ridges, deep-sea hydrothermal vents, subduction zones, and continental settings. Methane can also be generated by metamorphic reactions, particularly in subduction zones and continental settings such as ophiolites, orogenic massifs, and Precambrian shields. Both water–rock and metamorphic reactions can generate variable quantities of methane depending on the geochemical conditions, but, on Earth, methane fluxes are orders of magnitude smaller than biological sources. Finally, impacts or other exogenous sources can generate methane. The impact flux was larger during earlier periods in Earth's history, and such large impact fluxes are necessary to generate significant methane. A critical factor that influences the amount of methane that can be generated via all of these processes is the source of reducing power; in comparatively oxidizing surface environments with abundant $CO_2$, a reductant is needed to reduce carbon to $CH_4$. For magmatic outgassing, the reducing power ultimately comes from the mantle, with more reduced mantles outgassing more methane relative to $CO_2$ and CO. For low-temperature water–rock and metamorphic reactions, the key source of reducing power is ferrous iron ($Fe^{2+}$) in the crust, and in some cases the redox state of the mantle can also influence methane generation. For impact events, the metallic or ferrous iron that is delivered by the impactor serves as the source of reducing power.

2) low-temperature water–rock and metamorphic reactions, and 3) impact events.

**Volcanism/high-temperature magmatic outgassing.** Volcanoes on Earth today do not outgas significant methane. Most subaerial volcanoes produce less than $\sim 10^{-6}$ Tmol $CH_4$ per year (10, 58), and given the $\sim 1,500$ active volcanoes on Earth today, the estimated global $CH_4$ flux is $<10^{-3}$ Tmol/y, much less than the current biogenic flux of 30 Tmol/y. Similarly, Schindler and Kasting (6) estimated the $CH_4$ flux from submarine volcanism to be $\sim 10^{-2}$ Tmol/y. Although mud volcanoes, geological structures that transport clay rocks and sediment from Earth's interior to the surface, can emit large amounts of methane and $CO_2$ (59), the methane is largely thermogenic, ultimately deriving from organic matter produced by life (60). In principle, a terrestrial planet could abiotically emit methane through mud volcanoes given an abiotic source for the organic matter, such as hydrocarbon deposition from an organic haze. But that organic matter would need to be continuously replenished, and it is challenging for abiotic sources to provide the necessary replenishment (16, 42), especially under conditions sufficiently oxidizing to maintain a $CO_2$-rich atmosphere.

Wogan et al. (11) investigated whether magmatic outgassing could produce genuinely abiotic $CH_4$ fluxes on terrestrial planets with diverse compositions and surface conditions. They determined that volcanoes are unlikely to produce $CH_4$ fluxes comparable to Earth's biological flux because water has a high solubility in magma, which limits how much hydrogen (and therefore $CH_4$) can outgas. Also, $CH_4$ formation is thermodynamically favorable at temperatures lower than typical magma temperatures on Earth and at magma oxygen fugacities much more reduced than those expected for most terrestrial planets (11).

Could planets with significantly more reduced mantles and crusts produce high $CH_4$ fluxes via magmatic outgassing? Mercury's silicate interior has a low oxygen fugacity of $\sim 5$ $\log_{10}$ units below the iron-wüstite (IW) redox buffer, and its crust is enriched in graphite, a crystalline form of carbon (61, 62). While Mercury's small size and proximity to the Sun preclude the retention of an atmosphere, if there are large terrestrial exoplanets with similarly reducing interiors, then it is important to determine whether magmatic outgassing could produce $CH_4$-rich atmospheres.

Following the melting and volatile partitioning methods used in ref. 63, we applied a batch melting model, which assumes a partial melt is in equilibrium with the source rock before it rises to the surface, to determine the partitioning of volatiles from the rock to the melt (*SI Appendix*, section 6B). We assume the partitioning of carbon between the melt and solid phases is controlled by oxygen fugacity-dependent graphite saturation. For the top ~10 km of crust (pressures from ~0 to 0.5 GPa and solidus temperatures from ~1,400 to 1,445 K), we ran a Monte Carlo simulation to explore a range of source rock $CO_2$ and $H_2O$ concentrations, melt fractions, and planetary melt production volumes with oxygen fugacities from IW−11 to IW+5 (*SI Appendix*, Table S1). We find that for very reduced melts at or below IW−2, essentially all of the carbon (>99%) will precipitate as graphite during partial melting, so there is negligible carbon available for gaseous phases (Fig. 3 and *SI Appendix*, Fig. S2), consistent with observations of Mercury's graphite-enriched crust (64). Rocky exoplanets with ultrareduced magma compositions are unlikely to outgas significant quantities of $CH_4$ due to graphite saturation, although more experiments are needed to confirm reduced magmas' outgassing compositions.

In the rare cases where volcanoes could produce biogenic levels of $CH_4$ assuming magma production rates larger (>10 times) than those on Earth today, they would also outgas significant amounts of carbon monoxide (CO) gas (11). As described above, the atmospheric $CO/CH_4$ ratio could be used to distinguish between abiotic (outgassed) and biotic scenarios (11, 37). Ultimately, high-temperature magmatic outgassing, such as through volcanism, is unlikely to produce atmospheric $CH_4$ fluxes similar to those produced by biology on Earth.

***Low-temperature water–rock reactions and metamorphic reactions.*** The reliability of methane as a biosignature on habitable planets depends upon the tendency of low-temperature (below solidus) systems to generate methane via abiotic reactions. Under oxidizing planetary conditions conducive to $CO_2$ degassing, low-temperature $CH_4$ production is ultimately limited by the supply of reducing power in the form of ferrous iron ($Fe^{2+}$) in newly produced crust. One of the most frequently discussed processes for methane production is serpentinization, through which iron-bearing minerals are altered by hydration to produce $H_2$ via the oxidation of $Fe^{2+}$ by water (10, 69, 70):

$$3FeO + H_2O \longrightarrow Fe_3O_4 + H_2. \qquad [7]$$

Subsequently, $H_2$ can react with oxidized forms of carbon to produce $CH_4$ by Fischer–Tropsch-type (FTT) reactions:

$$4H_2 + CO_2 \longrightarrow CH_4 + 2H_2O. \qquad [8]$$

Metamorphic reactions may also produce $CH_4$ via iron oxidation. For example, Fe-bearing carbonates can decompose when metamorphosed and react with water to form $CH_4$ (71):

$$3\,FeCO_3 + H_2O \longrightarrow Fe_3O_4 + CO_2 + CO + CH_4$$
$$+ \text{Hydrocarbons}. \qquad [9]$$

Experimental methane and hydrocarbon yields via such reactions are typically very low compared to that of $CO_2$ (72).

Experimental, observational, and theoretical approaches have been taken to determine the efficiency of hydrothermal and metamorphic processes and their corresponding abiotic $CH_4$ production fluxes on Earth and how they may apply in other planetary environments. Various geological settings are potentially conducive to $CH_4$ generation, including midocean ridges, subduction zones, and continental settings. For example, Keir (73) and Cannat et al. (74) investigated the concentrations of $CH_4$ produced by serpentinization at midocean ridges and both found global abiotic $CH_4$ fluxes to be about three orders of magnitude smaller than the global biogenic $CH_4$ flux. Combining observational and theoretical approaches, Catling and Kasting (75) estimated abiotic hydrothermal $CH_4$ fluxes from both axial and off-axis vents ranging from 0.015 to 0.03 Tmol/y. In addition, Guzmán-Marmolejo et al. (7) and Kasting (8) determined abiotic $CH_4$ fluxes from hydrothermal systems ranging from 0.1 to 0.4 Tmol/y at present, and Kasting (8) found that this flux may potentially have been larger during the Hadean, ~1.5 Tmol/y, but this is still over an order of magnitude smaller than the current biogenic flux. Brovarone et al. (76) and Fiebig et al. (77) estimated abiotic hydrothermal $CH_4$ fluxes at subduction zones, finding modern fluxes of ~$10^{-2}$ Tmol/y similar to the above estimates. In continental settings, abiotic methane has been reported in low-temperature environments such as orogenic massifs and intrusions, seeps, crystalline shields, and ophiolites, with serpentinization of (iron-bearing) peridotites being the major source of methane in these settings (Fig. 2) (78). However, the amount of abiotic methane generated in continental settings is several orders of magnitude smaller than the biogenic flux (78–82).

Experimental studies on abiotic $CH_4$ production via water–rock and metamorphic reactions have also been conducted. The availability of $H_2$, the amount of excess aqueous carbonates, and the presence of mineral catalysts can greatly affect the amount of $CH_4$ generated experimentally (83, 84). While Oze et al. (84) and Neubeck et al. (85) found that $CH_4$ production by serpentinization is enhanced by the presence of mineral catalysts (e.g., chromite, magnetite, and awaruite), McCollom (71) cautions that these experimental studies did not quantify their organic contamination. McCollom (86) used isotopic labeling to differentiate $CH_4$ produced by serpentinization from background sources. McCollom (86) found abiotic $CH_4$ formation via serpentinization to be extremely limited, with most of the experimentally generated $CH_4$ deriving from background sources. While iron oxidation and FTT-type reactions (or their metamorphic equivalents) are the most commonly discussed mechanisms for large abiotic fluxes on terrestrial planets, other possible mechanisms for reducing carbon include direct carbonate methanation and hydration of graphite-carbonate–bearing rocks, but they are also unlikely to generate false-positive scenarios (*SI Appendix*, section 2).

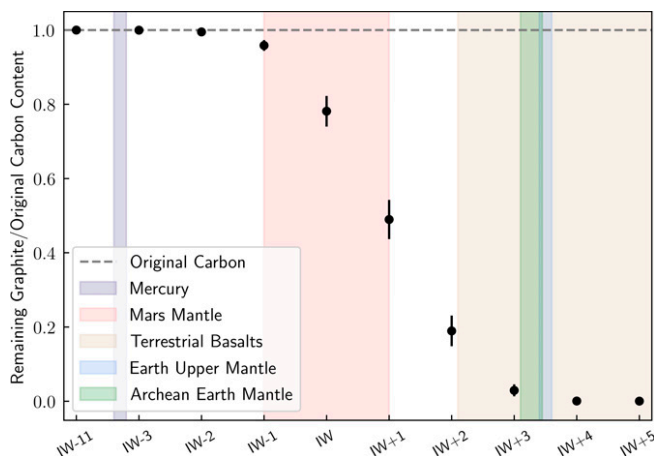

**Fig. 3.** Most carbon partitions into graphite under reducing conditions and so cannot degas as $CH_4$. Shown is the ratio of the amount of remaining graphite to the original carbon content as a function of oxygen fugacity. We used a batch-melting model to determine how volatiles would partition between the rock and melt over an ~10-km deep column of newly produced crust with pressures from ~0 to 0.5 GPa and temperatures from 1,400 to 1,445 K (*SI Appendix*, section 6B). For each oxygen fugacity, we ran a Monte Carlo simulation varying the input parameters, including $CO_2$ and $H_2O$ mass fractions in the mantle source rock, the fraction of source material that is melted during emplacement, and the planetary melt production rate. The average ratio of remaining graphite to initial carbon content from the Monte Carlo simulation is shown with the uncertainty reported as the 95% confidence interval. The horizontal dashed line (y = 1) illustrates the original amount of carbon, and ratios that fall on this line have all of the original carbon stable as graphite. The shaded vertical regions show the estimated oxygen fugacities of Mercury's lavas (61), the Martian mantle (65), terrestrial basalts (66), Earth's upper mantle (67), and Archean Earth's mantle (68) for reference.

The critical limitation of hydrothermal $CH_4$ production is the supply of $Fe^{2+}$ and the efficiency with which iron can be oxidized to generate $CH_4$. The availability of iron and the efficiency of its oxidation on a planetary scale depend on a range of geological and geochemical processes that operate across disparate spatial and temporal scales. Tectonic regime, mineral catalysis, volatile inventories, surface climate, and crustal composition and permeability/porosity all potentially modulate the efficiency and extent of crustal hydration. To investigate this process's limitations, Krissansen-Totton et al. (14) estimated the maximum $CH_4$ flux generated via serpentinization by exploring plausible ranges of parameters including crustal production rate, the fraction of FeO in fresh crust, the maximum fractional conversion of FeO to $H_2$ via serpentinization, and the maximum fractional conversion of $H_2$ to $CH_4$ via FTT reactions. Producing a probability distribution for the maximum abiotic $CH_4$ flux, they found that Earth-like biological $CH_4$ fluxes are at least an order of magnitude larger than plausible abiotic fluxes from serpentinization, consistent with the findings of the studies discussed above (14) (Fig. 4).

Ultimately, abiotic $CH_4$ generation via low-temperature water–rock or metamorphic reactions is unlikely to produce atmospheric $CH_4$ fluxes comparable to modern biotic fluxes in combination with atmospheric $CO_2$ (*SI Appendix*, Table S2 and Fig. 4). In fact, all $CH_4$ flux extrapolations from low-temperature system studies discussed above are consistent with the maximum abiotic flux estimates in ref. 14. Nevertheless, the possible parameter space for crustal methane production is vast, and work remains to be done to determine whether unfamiliar environmental conditions may exist on other planets that could produce a false-positive signal. For example, Fe-enriched olivine may be more common compositions for the mantles of other rocky planets compared to the Mg-rich olivine characteristic of Earth's mantle. McCollom et al. (87) determined that serpentinization of Fe-rich olivine can generate significantly more $H_2$ compared to that of Mg-rich olivine (by a factor of ~2 to 10) (87). Another source of uncertainty is what catalysts might be available in natural settings. At temperatures ≤600 K, in gas mixtures with $CO_2$ and $H_2$, $CH_4$ is thermodynamically preferred, but the reaction is kinetically inhibited and will proceed only if catalyzed. Future investigations could seek to develop coupled geochemical evolution models of a planet's mantle and crust that can self-consistently predict $CH_4$, $CO_2$, and CO fluxes from high-temperature magmatic processes and low-temperature hydrothermal and metamorphic systems, such that the contextual clues of abiotic methane can be explored for different compositional assumptions.

**Impacts.** The solar system terrestrial planets likely experienced a late-accreting veneer from impacts of comets and asteroids prior to 3.8 Ga (88). Impact events are plausible abiotic sources that can generate methane in two ways: 1) After a cometary impactor hits a planet, it vaporizes, and in the cooling impactor, some of the molecules delivered by the impactor may react to form $CH_4$ (89); and 2) large asteroid impactors could deliver a reducing power (i.e., iron) and vaporize a planet's surface ocean, causing a steam atmosphere to form, and $CH_4$ may form in such a cooling steam atmosphere (41). To generate significant methane, impact events require either a large, constant flux of impactors (case 1) or a transient postimpact atmosphere from a giant impact event (case 2).

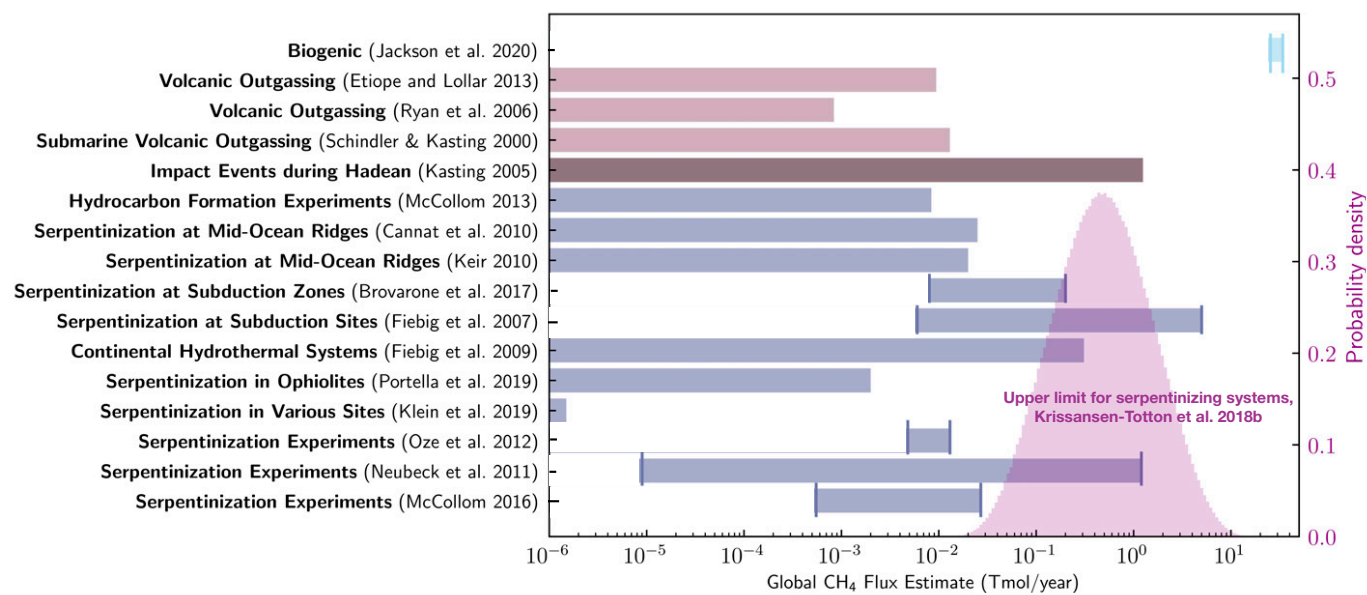

**Fig. 4.** Summary of known abiotic $CH_4$ sources with their estimated global $CH_4$ flux values compared to Earth's current biogenic $CH_4$ flux. As in *SI Appendix*, Table S2, for each abiotic source considered, we present those sources for which we can estimate global $CH_4$ flux values from a given reference. In the cases where there are multiple global $CH_4$ flux estimates for a given reference of an abiotic source, we show the maximum and minimum $CH_4$ flux estimates by the vertical lines (6, 8, 10, 14, 28, 58, 71, 73, 74, 76, 77, 79, 81, 82, 84–86). The transparent purple probability distribution for the maximum abiotic $CH_4$ flux from serpentinization is from ref. 14, and the right-hand $y$ axis shows the probability density of this distribution. None of the abiotic sources considered have estimated global $CH_4$ fluxes that are similar to or exceed Earth's modern biogenic $CH_4$ flux. In fact, most of the abiotic sources have predicted global $CH_4$ fluxes that are at least an order of magnitude less than Earth's biogenic $CH_4$ flux. We do not show the flux estimates that exceed the iron supply because such extremely large fluxes are based on experimental results for which there are issues with organic contamination (main text).

For case 1, Kress and McKay (89) and Kasting (8) modeled $CH_4$ formation from volatile-rich impactors. Ref. 89 found that a 1-km comet can generate 0.6 Tmol of atmospheric $CH_4$ per impact event, and ref. 8 estimated that the global $CH_4$ impact flux during the Hadean was ∼1.25 Tmol/y. However, it is unknown whether condensing dust from cometary impactors has effective catalytic properties to enable $CH_4$ generation. Recent theoretical and experimental work investigated the outgassing compositions of chondritic materials that may represent cometary impactors and found that there are small to negligible amounts of outgassed $CH_4$ from some of the most volatile-rich chondrites (i.e., CM chondrites) (90, 91).

For case 2, Zahnle et al. (41) showed that a transient reducing atmosphere (rich in $CH_4$, $H_2$, and $NH_3$) could have been generated on the early Earth by large asteroid impacts during the late-accreting veneer. Such giant impacts would produce methane since they delivered metallic iron, a significant reducing power, to the surface (41). The iron could react with Earth's existing $H_2O$ to produce $H_2$ and FeO, which would subsequently react with atmospheric $CO_2$ or CO to produce $CH_4$. The amount of methane that could form depends on the amount of carbon available prior to the impact, how much iron the impactor delivers, how much of that iron reacts with the atmosphere, and the presence of catalysts that can reduce the quench temperature so methane is thermodynamically stable (41). A possible false-positive scenario is one in which a giant impact event could produce a transient atmosphere with abundant $CH_4$ and $CO_2$ but low CO. However, calculations of transient impact-generated atmospheres of ref. 41 suggest that such false-positive scenarios are unlikely to be long lived for significant portions of geologic time and would be accompanied by $H_2$-dominated atmospheres (e.g., figures 7, 8, and 12 in ref. 41).

**Methane Beyond Earth: Mars and Temperate Exo-Titans.** Methane exists in other locations besides Earth throughout the solar system, including in the atmospheres of the outer planets and in comets (92). While super-Earths and sub-Neptune planets do not exist in our solar system, they are common among other planetary systems, and future studies could determine the surface pressures necessary for these planets to sustain methane via thermochemical recombination, without the need for a significant surface flux (*SI Appendix*, section 5). For example, if atmospheric $H_2$ is abundant, then $CH_4$ will efficiently recombine after photolysis, which dramatically increases the $CH_4$ lifetime (*SI Appendix*, section 3). As the focus of this study is on terrestrial planets, this section discusses atmospheric methane sources in other terrestrial worlds, in particular Mars and temperate Titan-like exoplanets (exo-Titans).

*Mars.* The presence of methane on Mars is debated, with claims of detections at the ∼10 to 60 ppbv level that are highly variable in time and space by the European Space Agency's (ESA) Mars Express, NASA's Curiosity rover, and ground-based observations (52, 93, 94, 95). However, the most recent and most sensitive measurements by the ESA-Roscosmos ExoMars Trace Gas Orbiter did not detect any significant methane over all observed latitudes and reported an upper limit of ∼20 ppt methane for altitudes above a few kilometers, several orders of magnitude lower than all previous purported $CH_4$ detections (96). Regardless, methane detections of a few parts per billion to tens of parts per billion are much lower than the terrestrial exoplanet thresholds for biogenic $CH_4$ considered in this study. There are a variety of plausible abiotic explanations for methane on Mars, including water–rock reactions, the release of clathrates, and degradation of organic matter.

*Temperate exo-Titans.* Methane exists (at ∼1 to 5%) in the $N_2$-rich atmosphere of Saturn's largest moon Titan (97). Photochemical models predict that the current $CH_4$ in Titan's atmosphere would be destroyed in ∼30 My unless there is a mechanism that resupplies $CH_4$ to the atmosphere (98, 99). Possible mechanisms for Titan's $CH_4$ resupply include its subsurface ocean, $CH_4$ clathrate hydrates in the crust, liquid hydrocarbons in the subsurface, or outgassing from the interior (100). While life has been suggested as a possible explanation (101), the absence of conventionally habitable surface conditions makes geochemical processes more attractive explanations.

Whatever the source of Titan's methane, temperate Titan-like exoplanets are unlikely to produce a $CH_4$ + $CO_2$ biosignature false positive. We estimate the atmospheric $CH_4$ lifetime for an Earth-sized exoplanet with a Titan-like volatile inventory that migrates to the habitable zone where all surface ice melts (see *SI Appendix*, section 6D for a scenario where ice remains). Given initial $CH_4$ and $CO_2$ reservoirs relative to $H_2O$ based on Titan's volatile inventory (102), we neglect oxidation via OH to be conservative and calculate the loss of $CH_4$ via diffusion-limited hydrogen escape (103). We assume that the atmospheric mixing ratio of $CH_4$ is 10%, which is conservative given the respective solubilities of $CH_4$ and $CO_2$ and plausible background $N_2$ inventories (*SI Appendix*, section 6D). We find that for planets with water mass fractions that are <1.0 wt% of the planet's mass, the atmospheric $CH_4$ lifetime is short at habitable-zone separations (less than ∼10 My) (Fig. 5). If the water mass fraction is ∼10 wt% of the planet's mass, then atmospheric $CH_4$ may last for longer periods of time (∼100 My), but even so the duration is much shorter than typical stellar ages. In any case, it will likely be possible to identify planets with such large water inventories via their low densities. Whether hydrogen's removal timescale could be dramatically lengthened via low loss rates or other large hydrogen reservoirs (while maintaining a $CO_2$-rich atmosphere) is a promising topic for future computational studies.

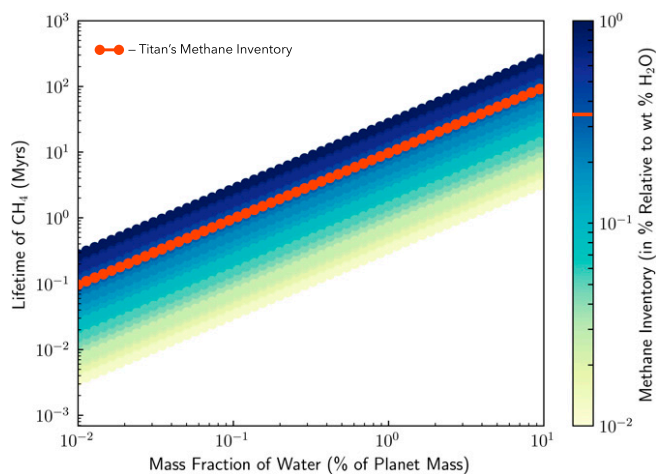

**Fig. 5.** The photochemical lifetime of methane biosignature false positives produced by melting volatile-rich Titan analogs is short. Shown is the estimated lifetime of atmospheric methane as a function of the planet's water mass and initial methane volatile inventory. Assuming methane's escape rate is diffusion limited and that its steady-state mixing ratio is 10%, we varied the initial methane volatile inventory (drawing values from a uniform distribution from 0.01 to 1.0% relative to weight % water, represented by the color bar) and the mass fraction of the planet's water (exploring values from 0.01 to 10% of the mass of the planet, assuming an Earth-mass planet) and calculated the estimated lifetime for methane in the atmosphere (*SI Appendix*, section 6D). The red curve represents Titan's methane inventory (∼0.35%) (102). For planets with Titan-like methane inventories and water mass fractions that are 1% (10%) of the planet's mass, the lifetime of atmospheric methane will be ∼10 My (∼100 My).

## Discussion

**Toward Procedures to Identify Methane Biosignatures.** Any procedure for observationally identifying methane biosignatures must take into account the broader planetary and astrophysical context and will be dictated by the capabilities of the available instruments. Major steps might include the following: 1) detecting a terrestrial planet within the habitable zone of its host star and characterizing its bulk properties (e.g., mass, radius, orbital properties); 2) measuring its atmospheric composition, namely the abundances of $CH_4$, $CO_2$, CO, $H_2O$, and $H_2$ and confirming that the atmosphere is anoxic; and 3) identifying possible false positives and combining this information with observational data on the planet's broader context to determine the likelihood of abiotic vs. biotic sources of methane (*SI Appendix*, Fig. S3). It is important that the host star is well characterized (i.e., UV radiation and stellar activity) to understand the planet's photochemical environment. Identifying the presence of liquid water on the surface of a planet would suggest a particularly compelling target since it is a likely requirement for life.

Constraining the atmospheric abundances of $CH_4$, $CO_2$, and CO and confirming that the atmosphere is not $H_2$ dominated is essential for determining whether the planet's atmosphere is indicative of the presence of a biosphere. Terrestrial planets with high mean-molecular-weight atmospheres are better candidates to search for methane biosignatures because in such atmospheres, the $CH_4$ lifetime will be very short without a significant replenishment source. In addition, confirming that the planet's atmosphere is anoxic is necessary to distinguish a false-positive case for an anoxic planet with abundant atmospheric $CH_4$, $CO_2$, and CO from an oxic planet with an oxygen-based biosphere that has atmospheric $CH_4$, $CO_2$, CO, and $O_2$ (37). With these abundances constrained, a photochemical model can infer the surface fluxes of the atmospheric constituents. Indications that these surface fluxes may be consistent with a biosphere include large implied $CH_4$ fluxes coexisting with atmospheric $CO_2$ but comparatively low CO abundances.

Even if the surface fluxes are consistent with a biosphere, it is necessary to identify all possible false positives including magmatic outgassing from a reduced mantle (Fig. 3), water–rock and metamorphic reactions (Fig. 4), large impact fluxes, and large volatile inventories (Fig. 5). The viability of detecting methane biosignatures depends on our knowledge of abiotic methane sources and their production rates. One of the most outstanding uncertainties is an incomplete understanding of plausible abiotic methane production on a planetary scale via water–rock and metamorphic reactions. If a planet has an atmospheric composition consistent with a methanogenic biosphere but false positives cannot be entirely ruled out, it will be necessary to search for corroborating evidence such as additional biosignature gases [e.g., methyl chloride (46), organosulfur compounds (104)], signs of atmospheric seasonality, and reflectance signatures from pigmented surface organisms (105, 106) (*SI Appendix*, Fig. S3). Ultimately, definitively detecting the presence of methane biosignatures on a terrestrial exoplanet will require taking into account the entire planetary and astrophysical context, characterizing the planet's atmospheric composition, investigating all potential false-positive scenarios, and likely searching for supporting evidence.

**Detectability Prospects.** Prospects for detecting biogenic levels of methane in terrestrial exoplanet atmospheres in the near future with JWST are promising (17, 24, 25, 27). However, it may be challenging to obtain sufficient observational data on the planetary context to confirm the presence of methane biosignatures and rule out false positives. Although JWST may be able to detect $CO_2$, it will provide only crude constraints on CO abundances (17, 27). Ref. 27 determined that JWST could place upper bounds on CO abundances in ~10 transits and constrain the $CO/CH_4$ ratio with more transits for an Archean Earth-like TRAPPIST-1e (27). Ref. 17 confirms that JWST will likely be able to crudely constrain the $CO/CH_4$ ratio and notes that CO constraints will be possible with high-resolution spectroscopy measurements with extremely large telescopes (ELTs). If biospheres are dominated by oxygenic photosynthesis, they may produce large CO fluxes through biomass burning (37). Therefore, to distinguish an anoxic, lifeless world with abundant atmospheric $CH_4$, $CO_2$, and CO from an oxic, inhabited planet with $CH_4$, $CO_2$, CO, and $O_2$ requires observations that can detect or rule out the presence of atmospheric $O_2/O_3$, which will be challenging with JWST (37). In addition, JWST will not be able to detect water vapor with transit observations due to water cloud condensation nor constrain surface properties, so it will not be able to fully assess habitability (107, 108). Nevertheless, if JWST detects significant $CH_4$ and $CO_2$ and places some constraints on the $CO/CH_4$ ratio in a terrestrial exoplanet's atmosphere, such a discovery would certainly motivate observations with future instruments.

Looking ahead, ground-based ELTs will help characterize terrestrial exoplanets and their biosignatures (109). Ref. 26 determined that for a cloud-free, low-$CO_2$ TRAPPIST-1e atmosphere, a mere 10 ppm $CH_4$ is likely detectable with high-resolution transit spectroscopy with the European ELT in less than ~30 transits, and CO detections may be possible with ~40 transits (26). In addition, the Astro2020 Decadal Survey recommended an ~6m infrared/optical/UV space telescope to characterize the atmospheres of dozens of habitable-zone terrestrial exoplanets, including detecting methane (5, 110). Identifying methane biosignatures will require not only detecting and constraining the atmospheric abundances of $CH_4$, $CO_2$, and CO, but also using a combination of observational tools to comprehensively characterize the broader planetary context.

## Conclusions

With the upcoming technological advancements in exoplanet observations enabling the characterization of potentially habitable exoplanets, it is important to consider possible biosignature gases and the sources of false-positive detections. This is particularly urgent for methane since biogenic methane is likely detectable for some terrestrial exoplanets with JWST. The case for methane as a biosignature stems from the fact that photochemistry of terrestrial planet atmospheres implies that large $CH_4$ surface fluxes are required to sustain high levels of atmospheric methane. Although a variety of abiotic mechanisms could, under diverse planetary environments, replenish atmospheric methane, we find that it is challenging for such sources to produce abiotic $CH_4$ fluxes comparable to Earth's biogenic flux without also generating observable contextual clues that would signify a false positive. For example, we investigated whether planets with very reduced mantles and crusts can generate large methane fluxes via magmatic outgassing and assessed the existing literature on low-temperature water–rock and metamorphic reactions and, where possible, determined their maximum global abiotic methane fluxes. In every case, abiotic processes cannot easily produce atmospheres rich in both $CH_4$ and $CO_2$ with negligible CO due to the strong redox disequilibrium between $CO_2$ and $CH_4$ and the fact that CO is expected to be readily consumed by life. We also explored whether habitable-zone exoplanets that have large volatile inventories like

Titan could have long lifetimes of atmospheric methane. We found that, for Earth-mass planets with water mass fractions that are less than ∼1% of the planet's mass, the lifetime of atmospheric methane is less than ∼10 My, and observational tools can likely distinguish planets with larger water mass fractions from those with terrestrial densities.

Clearly, the mere detection of methane in an exoplanet's atmosphere is not sufficient evidence to indicate the presence of life given the variety of abiotic methane-production mechanisms. Instead, the entire planetary and astrophysical context must be taken into account to interpret atmospheric methane. *SI Appendix*, Fig. S3 illustrates a tentative procedure for identifying methane biosignatures in the atmospheres of habitable terrestrial exoplanets. Ultimately, methane is more likely to be biogenic on a habitable-zone planet when 1) planet bulk density is terrestrial (no large surface volatile reservoirs), the atmosphere has a high mean molecular weight and is anoxic, and the host star is old; 2) the atmospheric $CH_4$ abundance is high, with implied surface replenishment fluxes exceeding what could plausibly be produced by known abiotic processes (∼10 Tmol/y); and 3) when atmospheric methane is accompanied by $CO_2$ but comparatively little CO (or $CO/CH_4 < 1$).

## Materials and Methods

We use the photochemical model PhotochemPy in *SI Appendix*, Fig. S1 (*SI Appendix*, section 6A). The calculations for determining how carbon partitions between different phases under various redox conditions for Fig. 3 follow the methods in ref. 63 and are discussed further in *SI Appendix*, section 6B. The global abiotic $CH_4$ flux estimates in Fig. 4 are described in detail in *SI Appendix*, section 6C. For Fig. 5, we estimate the atmospheric $CH_4$ lifetime for an Earth-mass terrestrial planet with different water mass fractions and Titan-like volatile inventories by assuming the escape flux of hydrogen is diffusion limited (*SI Appendix*, section 6D). The codes used for our analysis are available on GitHub at https://github.com/maggieapril3/MethaneBiosignature (*SI Appendix*, section 6).

**Data Availability.** All data needed to evaluate the conclusions in this paper are present in this paper and/or in *SI Appendix*, Materials and Methods. PhotochemPy can be accessed at GitHub (https://github.com/Nicholaswogan/PhotochemPy). Python code data have been deposited in GitHub (https://github.com/maggieapril3/MethaneBiosignature) (111).

**ACKNOWLEDGMENTS.** We thank James Kasting and the other anonymous reviewer for constructive reviews. We thank David Catling, Edward Schwieterman, Xinting Yu, Kevin Zahnle, and Stephanie Olson for helpful discussions and comments. We thank Elena Hartley (http://www.elabarts.com) for creating Fig. 2. J.K.-T. is supported by the NASA Sagan Fellowship and through the NASA Hubble Fellowship Grant HF2-51437 awarded by the Space Telescope Science Institute, which is operated by the Association of Universities for Research in Astronomy, Inc., for NASA, under Contract NAS5-26555. N.W. is supported by the NASA Astrobiology Program Grant 80NSSC18K0829. M.T. is supported by NASA Emerging Worlds Grant 80NSSC18K0498 and NASA Planetary Science Early Career Award Grant 80NSSC20K1078. M.A.T., M.T., and J.F.F. are supported by NASA under Award 19-ICAR19_2-0041.

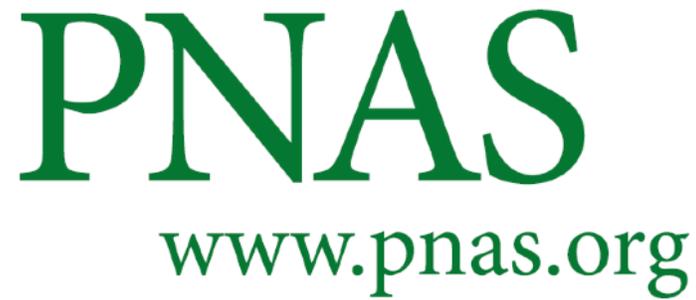

# Supplementary Information for

**The Case and Context for Atmospheric Methane as an Exoplanet Biosignature**

**Maggie A. Thompson, Joshua Krissansen-Totton, Nicholas Wogan, Myriam Telus, Jonathan J. Fortney**

**Maggie A. Thompson.**
**E-mail: maapthom@ucsc.edu**

**This PDF file includes:**

Supplementary text
Figs. S1 to S3 (not allowed for Brief Reports)
Tables S1 to S3 (not allowed for Brief Reports)
SI References

## Supporting Information Text

### 1. Atmospheres with Abundant CH$_4$ and CO$_2$ in Chemical Equilibrium–Discussion of Woitke et al. 2021

Could there be a biosignature false-positive scenario in which abundant CH$_4$ and CO$_2$ (without CO) coexist in thermochemical equilibrium in a rocky planet's atmosphere? Woitke et al. 2021 (1) show that at temperatures below ∼600 K, CO$_2$, CH$_4$, N$_2$, and H$_2$O could coexist in chemical equilibrium if aqueous species are neglected. However, Woitke et al. 2021 (1) did not consider photochemistry and its effects on the stability of these atmospheres. Photochemistry is essential when assessing the plausibility of proposed terrestrial planet atmospheres.

To illustrate how a large CH$_4$ surface flux would be required to sustain high levels of atmospheric CH$_4$, a series of photochemical models were run simulating terrestrial planet atmospheres. We used PhotochemPy, a photochemical model adapted from the Atmos code (2) and created by N. Wogan ((3), https://github.com/Nicholaswogan/PhotochemPy), that uses a set of inputs (e.g., stellar flux, atmospheric temperature structure, chemical species and reactions) and then integrates the atmosphere forward in time until it reaches a photochemical steady state (see SI Section 6A). A series of models were generated assuming a planet with an initial atmospheric composition that is Archean Earth-like (i.e., N$_2$-CO$_2$-CH$_4$), orbiting a 2.7 Ga Sun-like star and explored a range of CO$_2$ and CH$_4$ surface mixing ratios from 0.1 to 0.5 and from $10^{-5}$ to 0.1, respectively.

Figure S1 illustrates that abundant atmospheric CH$_4$ in atmospheres containing CO$_2$, N$_2$, and H$_2$O, requires CH$_4$ surface fluxes similar to or greater than modern Earth's biogenic flux to balance photochemical destruction. Therefore, such an atmosphere will not be stable without a significant CH$_4$ replenishment source that likely exceeds Earth's modern biogenic flux. For example, in order to sustain high atmospheric CH$_4$ mixing ratios of ∼0.1 along with significant amounts of CO$_2$, the required CH$_4$ surface fluxes are on the order of $\sim 3.7 \times 10^{12}$ molecules/cm$^2$/s (∼1000 Tmol/year), corresponding to the yellow regions of Figure S1. Such a large CH$_4$ replenishment flux would be about 30 times larger than Earth's current biogenic flux (30 Tmol/year). Considering the global redox budget, such abundant atmospheric CH$_4$ requires that either the flux of reductants from Earth's interior is at least three orders of magnitude higher than Earth's modern hydrogen outgassing rate or that the H$_2$ escape rate is much less than the diffusion limit (4). In addition, the equilibrium calculations of (1) did not consider the formation of dissolved ammonium (NH$_4^+$) and bicarbonate (HCO$_3^-$), which are shown to be the equilibrium products of CO$_2$, CH$_4$, and N$_2$ in the presence of liquid water in (5). The thermochemical calculations of (1) could instead have relevance for deep sub-Neptune atmospheres.

### 2. Additional Water-Rock and Metamorphic Reactions and Key Unknowns

While iron oxidation and FTT-type reactions (or their metamorphic equivalents) are the most commonly discussed mechanisms for large abiotic fluxes on terrestrial planets (Figure 2), it is worth considering other possible mechanisms for reducing carbon. Direct carbonate methanation can produce CH$_4$ given an exogenous supply of H$_2$ (6–9), as follows:

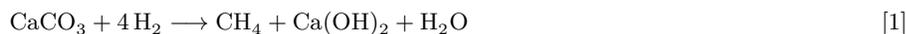
$$\mathrm{CaCO_3 + 4\,H_2 \longrightarrow CH_4 + Ca(OH)_2 + H_2O} \qquad [1]$$

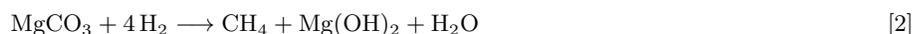
$$\mathrm{MgCO_3 + 4\,H_2 \longrightarrow CH_4 + Mg(OH)_2 + H_2O} \qquad [2]$$

Here, the production of H$_2$ is likely to be limited by iron oxidation via water rock reactions, unless conditions are sufficiently reducing that H$_2$ rather than H$_2$O is the dominant H-bearing product from magmatic outgassing. In this scenario, however, simultaneously large fluxes and atmospheric concentrations of CO$_2$ are unlikely (Figure S2).

Hydration of graphite-carbonate bearing rocks can similarly generate CH$_4$ without the need for iron oxidation (10), in the following reaction:

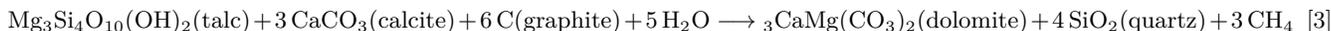
$$\mathrm{Mg_3Si_4O_{10}(OH)_2(talc) + 3\,CaCO_3(calcite) + 6\,C(graphite) + 5\,H_2O \longrightarrow {}_3CaMg(CO_3)_2(dolomite) + 4\,SiO_2(quartz) + 3\,CH_4} \quad [3]$$

However, in the absence of biological organic matter, crustal compositions rich in graphite, require strongly reducing conditions (Figure 3). These are unlikely to coexist with high magmatic CO$_2$ fluxes, which require oxidizing conditions, without large magmatic fluxes of CO (Figure S2).

Some of the other fundamental unknowns with regards to water-rock and metamorphic reactions include the efficiency and extent of hydration reactions under different tectonic regimes, the importance of carbonate hydration in the presence of graphite in generating CH$_4$, the extent to which H$_2$ can directly react with carbonates to produce CH$_4$ under reducing melt conditions, and the extent to which heterogeneous surface environments could simultaneously produce high CH$_4$ and CO$_2$ fluxes.

### 3. Photochemical Destruction and Recombination Pathways for Methane

Methane is removed from an atmosphere photochemically in two ways, depending on the concentration of CO$_2$ relative to CH$_4$ and the presence of other oxidants (11). In the first case where CO$_2$ is more abundant, then CH$_4$ is destroyed by oxidants and ultimately is converted to CO$_2$, such as through the following reactions:

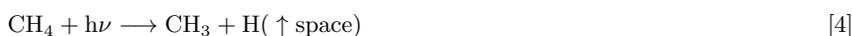
$$\mathrm{CH_4 + h\nu \longrightarrow CH_3 + H(\uparrow space)} \qquad [4]$$

or

$$CH_4 + OH \longrightarrow CH_3 + H_2O \quad [5]$$

or

$$CH_4 + O^1D \longrightarrow OH + CH_3 \quad [6]$$

and, subsequently, either

$$CH_3 + O \longrightarrow H_2CO + H(\uparrow space) \quad [7]$$

or

$$CH_3 + O_2 \longrightarrow H_2CO + OH \quad [8]$$

The C in $H_2CO$ is further oxidized to $CO_2$. The H produced can then be lost to space, thereby irreversibly destroying the $CH_4$. Note that OH, $O^1D$, O and $O_2$ are byproducts of $H_2O$ and $CO_2$ photolysis; an atmosphere rich in molecular oxygen is not required for rapid $CH_4$ destruction (although it does decrease the $CH_4$ lifetime). The pathway involving OH and $O_2$ is the dominant destruction pathway in Earth's modern atmosphere, and all of these pathways were likely important for the Archean atmosphere (11).

For the second case where $CH_4$ is more abundant than $CO_2$, $CH_4$ polymerizes to aerosols, which fall to the ground and remove the atmospheric $CH_4$. The chemistry producing aerosols is complex, but this sequence of reactions (Equations 9-14) demonstrates the general process:

$$CH_4 + h\nu \longrightarrow {}^1CH_2 + H_2(\uparrow space) \quad [9]$$

$${}^1CH_2 + N_2 \longrightarrow {}^3CH_2 + N_2 \quad [10]$$

$${}^3CH_2 + {}^3CH_2 \longrightarrow C_2H_2 + H_2(\uparrow space) \quad [11]$$

$$C_2H_2 + h\nu \longrightarrow C_2H + H(\uparrow space) \quad [12]$$

$$C_2H_2 + C_2H \longrightarrow C_4H_2 + H(\uparrow space) \quad [13]$$

$$C_4H_2 + C_2H \longrightarrow \text{bigger hydrocarbons} \quad [14]$$

These hydrocarbons condense into aerosols that fall to the ground and thus remove $CH_4$ from the atmosphere. These aerosols could break down and release $CH_4$ back into the atmosphere or they could get buried and subducted into the planet. However, some portion of the hydrogen produced by methane photolysis will be lost to space, and so, without $H_2$ replenishment, the C:H ratio of condensate material will rise such that the methane is irreversibly lost.

Ultimately, the lifetime of atmospheric $CH_4$ is determined by the efficiency of the pathways outlined above. Atmospheric composition is an important determinant of that efficiency. For example, if $H_2$ is abundant, then $CH_4$ will efficiently recombine after photolysis via:

$$CH_3 + H + M \longrightarrow CH_4 + M \quad [15]$$

where M is an unspecified collision partner that carries away excess energy, which dramatically increases the $CH_4$ lifetime.

## 4. Gas Giant Planets

The giant planets in the Solar System contain abundant methane in their $H_2$-rich atmospheres due to accreting and processing primordial material from the solar nebula (12, 13). Methane in the atmospheres of giant planets can be replenished indefinitely because, although methane gets photodissociated in the upper atmosphere, hydrogen is never depleted via escape, and carbon and hydrogen can recombine deeper in the atmosphere where temperatures and pressures are high enough for methane production to be thermodynamically favorable and kinetically viable (14). Conversely, temperate terrestrial planets with high mean molecular weight atmospheres and surfaces do not have deep enough atmospheres to replenish methane without an additional source (abiotic or biotic). In terrestrial atmospheres without a replenishment source, methane is photodissociated and hydrogen is lost to space on short timescales (e.g., ∼10's of thousands of years for ∼1 bar atmospheres). However, if a $H_2$-rich atmosphere on a terrestrial planet exists, then the $CH_4$ lifetime would be long due to stabilizing reaction pathways like those that operate in giant planet atmospheres.

## 5. Super-Earths and Sub-Neptune Planets

Although the focus of this study is on methane biosignatures on terrestrial planets, it is important to consider methane in the atmospheres of super-Earths (planets with radii between 1 and ∼1.8 $R_⊕$) and sub-Neptunes (radii larger than ∼1.8 $R_⊕$ and less than ∼3.5 $R_⊕$) (15). These planets are expected to span a diverse range in bulk compositions from rocky to gaseous, and therefore their atmospheres are also expected to have various compositions, from thick, $H_2$/He-rich primary atmospheres to thin secondary atmospheres comparable to that of terrestrial planets. Methane in sub-Neptune atmospheres would be unremarkable due to the possibility of thermodynamic recombination at depth. However, future studies could seek to determine the atmospheric pressure necessary for a planet to sustain methane via thermodynamic recombination against photodissociation. Observational methods to distinguish super-Earths from sub-Neptunes are also relevant for excluding deep atmosphere replenishment as a source of methane (16).

When searching for methane in the atmospheres of super-Earths and sub-Neptunes, as is the case for terrestrial planets, it is also important to understand the effects of the host star. For example, M dwarfs tend to be more active longer into their life cycles and have stronger far-UV and weaker near-UV emissions compared to solar-type stars, making it essential to determine how such host stars may impact such an exoplanet's atmosphere (e.g., photochemistry). Future work should aim to couple geochemical and photochemical models to better understand how cooler host stars can affect a planet's atmosphere.

## 6. Materials and Methods

All codes used in this study are publicly available at https://github.com/maggieapril3/MethaneBiosignature.

**A. Photochemical Model: PhotochemPy.** We use the PhotochemPy photochemical model in Figure S1 to illustrate the methane fluxes required to sustain atmospheric methane in various atmospheres. PhotochemPy is a descendant of the Photochem model contained in the Atmos modeling suite, which was originally developed by Jim Kasting and Kevin Zahnle (17) and has since been developed by many of their students and colleagues. Appendix B in (18) contains an in-depth description of the main equations solved in the Photochem model. The physics and chemistry in PhotochemPy are very similar to the physics and chemistry in the version of Atmos used in (2). The main exception is that we have updated several reactions, and water photolysis cross sections following (19). However, PhotochemPy deserves a distinct name because it is a modern Fortran rewrite of the original Fortran 77 code. Additionally, PhotochemPy uses Numpy F2PY (20) to generate a Python wrapper to the compiled Fortran library ((3), https://github.com/Nicholaswogan/PhotochemPy).

**B. Carbon Partitioning and Magmatic Outgassing Calculations.** For Figure 3 and Figure S2, to calculate how carbon partitions between different phases under various redox conditions for ∼10 km of crust (pressures from ∼0-0.5 GPa and solidus temperatures from ∼1400-1445 K), we follow the melting and volatile partitioning methods outlined in (21). We first compute batch melting with standard partition coefficients for $CO_2$ and $H_2O$ which returns how concentrated $CO_2$ and $H_2O$ are in the melt for a given melt fraction, F:

$$X_{CO_2}^{\text{melt}} = (1 - (1-F))^{1/0.002}(m_{\text{mantle},CO_2}/m_{\text{mantle}})/F \quad [16]$$

$$X_{H_2O}^{\text{melt}} = (1 - (1-F))^{1/0.01}(m_{\text{mantle},H_2O}/m_{\text{mantle}})/F \quad [17]$$

where $m_{\text{mantle}}$ is the mantle mass (in kg, which for Earth is $4 \times 10^{24}$ kg) and $m_{\text{mantle},CO_2}$ and $m_{\text{mantle},H_2O}$ are the masses of $CO_2$ and $H_2O$ in the mantle, respectively (in kg). However, if the source material is reducing, the melt will never get this carbon rich due to graphite saturation. Therefore, we compute the carbon melt concentration assuming graphite saturation. We assume the carbon is stored in the mantle as graphite and dissolves into the melt as carbonate ions ($CO_3^{2-}$). The amount of carbonates dissolved in the melt are calculated from equilibrium constants $K_1$ and $K_2$:

$$X_{CO_3^{2-}}^{\text{melt}} = \frac{K_1 K_2 f_{O_2}}{1 + K_1 K_2 f_{O_2}} \quad [18]$$

$$log_{10} K_1 = 40.07639 - 2.53932 \times 10^{-2} T + 5.27096 \times 10^{-6} T^2 + 0.0267\frac{(P-1)}{T} \quad [19]$$

$$log_{10} K_2 = -6.24763 - \frac{282.56}{T} - 0.119242\frac{(P-1000)}{T} \quad [20]$$

for temperature ($T$, in K) and pressure ($P$, in bars) and $f_{O_2}$, which is the oxygen fugacity calculated based on the Iron-Wüstite redox buffer:

$$f_{O_2} = 10^{-27215/T + 6.57 + 0.0552(P-1)/T} \quad [21]$$

Then, we calculate the $CO_2$ melt abundance:

$$X_{CO_2}^{\text{melt}} = [\frac{M_{CO_2}}{\text{fwm}} X_{CO_3^{2-}}^{\text{melt}}]/[1 - (1 - \frac{M_{CO_2}}{\text{fwm}})X_{CO_3^{2-}}^{\text{melt}}] \quad [22]$$

where $M_{CO_2}$ is CO$_2$'s molar mass and fwm is the formula weight of the melt (36.594) (21). We take the minimum of the graphite-saturated CO$_2$ melt concentration (Equation 22) and the constant partition coefficient concentration (Equation 16) and use that $X_{CO_2}^{\text{melt}}$ value to calculate the flux of CO$_2$ in the melt:

$$F_{CO_2}^{\text{melt}} = X_{CO_2}^{\text{melt}} m_{\text{melt}} F \qquad [23]$$

where $m_{\text{melt}}$ is the planetary melt production rate (in g/s, where Earth's nominal melt production rate is $3.2 \times 10^9$ g/s). To compare the original carbon content to the amount that remains as graphite, we compute:

$$\text{Original Carbon} = \frac{m_{\text{mantle},CO_2}}{m_{\text{mantle}}} m_{\text{melt}} \qquad [24]$$

$$\text{Remaining Graphite} = \text{Original Carbon} - F_{CO_2}^{\text{melt}} \qquad [25]$$

For the Monte Carlo simulation, we vary the input parameters and sample distributions according to Table S1.

To calculate the species that would be released by magmatic outgassing and their corresponding outgassing fluxes, we use the outgassing speciation model described in (22). As magma ascends to the surface, the overburden pressure decreases and the dissolved volatiles in the magma may reach saturation. At that point, volatiles exsolve from the magma and form gas bubbles, which can be released to the atmosphere. In addition, chemical reactions take place within the bubbles which can alter their chemical compositions. This model estimates the composition of these gas bubbles just prior to their release to the atmosphere by solving a system of equations including the Iacono-Marziano solubility relationships for H$_2$O and CO$_2$, gas-phase equilibrium relationships and mass conservation of hydrogen and carbon (22). The model assumes that the oxygen fugacity of the gas is set by the oxygen fugacity of the magma and requires the following inputs: initial concentrations of H$_2$O and CO$_2$ in the melt prior to outgassing, temperature and pressure of outgassing, and redox state of the melt. Please refer to (22) for further details.

Swain et al. 2021 (23) recently claimed the detection of a thin, reducing atmosphere around the rocky exoplanet GJ 1132 b that is H$_2$-dominated and rich in CH$_4$ (∼0.5%). They simulated mantle outgassing and argued that an ultra-reduced magma could reproduce the observed atmosphere, with the best fitting model parameters giving an extremely reduced oxygen fugacity of $log f_{O_2} = IW - 11$ (23). Such reduced conditions could conceivably arise if significant amounts of hydrogen from a planet's primary atmosphere dissolve into the magma ocean and were sequestered into the interior, thereby providing a reservoir of volatiles that can later be outgassed to form a secondary atmosphere (23, 24). However, the outgassing model of (23) omits carbon partitioning between solid, liquid and gas phases under ultra-reducing redox conditions.

We find that for an ultra-reduced melt of $log f_{O_2} = IW - 11$, essentially all of the carbon (>99%) will remain saturated as graphite during partial melting, so there is negligible carbon available for gaseous phases (Figure 3). To confirm this, we used the above outgassing speciation model which solves for the gas-gas and gas-melt equilibrium in a C-O-H system, to predict the gases that would be released from the melt by magmatic outgassing. From this model, we determine that the CH$_4$, CO$_2$, and CO outgassing fluxes would be negligible (<1E-10 Tmol/year) at such a reduced oxygen fugacity (Figure S2). It is important to note that these calculations do not include carbon in the form of iron carbonyls and methyl (CH$_3$) groups bonded to Si$^{4+}$ in the melt as some studies have suggested will be present under reducing conditions (25, 26). However, it is not expected that these additional carbon-bearing species will significantly alter our findings as these studies also found carbon stable in the melt under reducing conditions. Therefore our findings suggest that the outgassing mechanism proposed for GJ 1132 b is improbable. Additionally, an independent study that analyzed the same Hubble Space Telescope (HST) transit data found no methane signature and instead claim a featureless spectrum for GJ 1132 b (27), and the findings of (23) conflict with Lyman-alpha observations of the system (28).

**C. Calculations of Global CH$_4$ Flux Estimates from Abiotic Sources.** The global CH$_4$ flux estimates for different abiotic sources in Table S2 (and illustrated in the schematic Figure 2) and Figure 4 are calculated as follows:

**Higher Temperature:**

- **Etiope & Lollar 2013:** To determine an estimated global abiotic CH$_4$ flux from surface volcanism, we take their CH$_4$ flux estimates for individual volcanoes and multiply these values by the number of active volcanoes on Earth today (∼1500) (6).

- **Ryan et al. 2006:** To determine an estimated global abiotic CH$_4$ flux from surface volcanism, we followed the same procedure as described above for Etiope & Lollar 2013, except with Ryan et al.'s estimates for individual volcanic outgassing CH$_4$ fluxes (29).

- **Schindler & Kasting 2000:** They estimated the current global CH$_4$ flux from submarine volcanism to be $\sim 10^{-2}$ Tmol/year by taking an observed ratio of CH$_4$/CO$_2$ in mid-ocean ridge hydrothermal vent fluids and an estimated total outgassed carbon flux at the mid-ocean ridges.

**Lower Temperature:**
**Observational and Theoretical Studies:**

- **Cannat et al. 2010:** They derive a global serpentinization-related methane flux at slow spreading mid-ocean ridges of $2.5 \times 10^{-2}$ Tmol/year (30).

- **Keir 2010:** Keir calculates a global methane flux of $2 \times 10^{-2}$ Tmol/year from mid-ocean ridges, similar to the findings of Cannat et al. 2010 (31).

- **Jones et al. 2010:** Their serpentinization experiments for mid-ocean ridges and forearcs determined $CH_4$ production rates ranging from $1 \times 10^{-5}$ to 0.05 $\mu$mol/kg/hr (32). Taking these rates and extrapolating to the mass of the entire oceanic crust, we find that the corresponding global abiotic $CH_4$ estimates exceed the Earth's iron supply. However McCollom 2013 (33) caution that Jones et al. did not determine their background levels of $CH_4$ so it is possible that a portion of the methane generated in their experiments came from sources of contamination. Therefore, the findings of Jones et al. cannot be extrapolated to a global abiotic flux.

- **Catling & Kasting 2017:** Using a combination of different observational studies, they determined abiotic $CH_4$ flux estimates for hot, axial vents and off-axis vent fields of 0.015 and 0.03 Tmol/year, respectively (4). For hot, axial vents, they estimated the abiotic $CH_4$ flux by using observed $CH_4$ and $CO_2$ concentrations from East Pacific Rise fluids. For the off-axis vent fields, they used an estimated $H_2$ flux and the $CH_4/H_2$ ratio in ultramafic vent fields to determine an abiotic $CH_4$ flux estimate (4).

- **Guzmán-Marmolejo et al. 2013:** They estimated the amount of $CH_4$ generated by serpentinization in hydrothermal vent systems. For $1M_\oplus$ and 5 $M_\oplus$ planets, they determined abiotic $CH_4$ fluxes of 0.18 Tmol/year and 0.35 Tmol/year, respectively (34). These estimates account for the supply rate of available FeO in the crust which is determined in part by the crustal production rate. For the 5 $M_\oplus$ planet, the crustal production rate is scaled from Earth using a power law. They also take into account the limitations of $CO_2$ in hydrothermal systems based on different observational studies.

- **Kasting 2005:** Kasting 2005 estimated the global abiotic $CH_4$ flux from off-axis mid-ocean ridges by extrapolating from observed methane concentrations in hydrothermal fluids. They found that, at present, the abiotic hydrothermal $CH_4$ flux is $\sim$0.1 Tmol/year, but during the Hadean the flux may have been larger, $\sim$1.5 Tmol/year (35). They note that if seafloor creation during the Hadean was much faster, the abiotic $CH_4$ flux could have increased by a factor of 5-10. However, given that seafloor production rates during that period are uncertain, we only include their Hadean estimate of 1.5 Tmol/year in Figure 4 and Table S2. These estimates are about an order of magnitude larger than those of Catling & Kasting 2017 because Kasting 2005 assumed a larger water circulation rate compared to that of Catling & Kasting 2017 (4). In addition, the Kasting 2005 Hadean abiotic flux estimate is larger than Guzman-Marmolejo et al.'s estimates because Guzman-Marmolejo took into account the iron supply and $CO_2$ limitations in hydrothermal systems (34, 35).

- **Brovarone et al. 2017:** This study determined several different global $CH_4$ flux estimates for different sites where serpentinization takes place including subduction zone fluids, forearc mantle wedges above subduction zones, and sub-seafloor conditions (36).

- **Fiebig et al. 2007:** They investigated subduction-related hydrothermal sites in the Mediterranean and computed both an uppermost flux estimate for abiogenic $CH_4$ during the Archean (2.5-5 Tmol/year) and a present-day flux ($6 \times 10^{-3}$ Tmol/year) (37).

- **Fiebig et al. 2009:** This study estimated the abiogenic $CH_4$ flux from continental hydrothermal systems to be 0.31 Tmol/year (38).

- **Portella et al. 2019:** Their study of serpentinization of chromitites in ophiolites found that chromitites can contain $CH_4$ gas concentrations up to 0.31 $\mu$g/g$_{\text{rock}}$. Taking this $CH_4$ concentration and multiplying it by Earth's melt production rate ($3.2 \times 10^9$ g/s) results in a global $CH_4$ flux estimate of $2 \times 10^{-3}$ Tmol/year (39).

- **Klein et al. 2019:** They studied methane formation in olivine-hosted secondary fluid inclusions to inform serpentinization in subduction zones, mid-ocean ridges and ophiolites. They determined that the Chimaera serpentinization system has released 0.076 to 0.5 km$^3$ $CH_4$ during the past 2000 years which is equivalent to $2 \times 10^6$-$11 \times 10^6$ mol/year of $CH_4$. They also estimated that the lower oceanic crust contains a total of $\sim$300 Tmol of $CH_4$ gas (40). Given that the lifetime of the oceanic crust is $\sim$200 Myrs, the estimated global abiotic $CH_4$ flux due to serpentinization is $1.5 \times 10^{-6}$ Tmol/year.

**Experiments:**

- **McCollom 2013:** The hydrocarbon formation experiments discussed in McCollom 2013 measure the amount of dissolved $CH_4$ gas in serpentinized olivine at 300 $^o$C as a function of time (see their Figure 7). Taking their experimental methane production rate of 0.05 $\mu$mol/kg/hr and scaling it to the entire mass of oceanic crust on Earth ($\sim 6 \times 10^{21}$ kg) results in a global flux estimate that exceeds Earth's iron supply (33). Such an estimate requires the whole crust to be at a high temperature which is unrealistic for a habitable zone terrestrial planet. Therefore, it is not possible to extrapolate their experimental findings to a global abiotic flux rate.

- **Oze et al. 2012:** They performed experiments to investigate the influence of mineral catalysts on serpentinization and found that the $CH_4$ production rate varied from $\sim$0.09 to 0.15 $\mu$mol/kg/hr. If we take their experimental $CH_4$ production rates and extrapolate to the mass of the oceanic crust, we find a global $CH_4$ flux estimate that exceeds Earth's iron supply. As with Jones et al. (32), McCollom 2013 (33) note that Oze et al. did not quantify their background levels of $CH_4$, so it is not possible to properly extrapolate a global abiotic $CH_4$ flux from their experiments.

- **Neubeck et al. 2011:** Their serpentinization experiments on forsteritic olivine determined $CH_4$ accumulation rates ranging from $2.7 \times 10^{-11}$ to $7.3 \times 10^{-11}$ mol/m$^2$/s. However, McCollom 2013 (33) noted that Neubeck et al. did not quantify their background $CH_4$ levels, so it is not possible to extract an abiotic global flux estimate from this study.

- **McCollom 2016:** They performed serpentinization experiments with olivine and measured a range of dissolved $CH_4$ concentrations from 5.5 to 270 $\mu$mol/kg$_{\text{olivine}}$. They used isotopic labeling to differentiate $CH_4$ produced by serpentinization from background sources, and found that in almost all experiments, the majority of $CH_4$ produced actually derived from background sources rather than from reduction of dissolved inorganic carbon. Using the isotopic labeling, for the experiments performed at or above 300 $^o$C, the amount of $CH_4$ generated via reduction of inorganic carbon was 16-50 $\mu$mol/kg$_{\text{olivine}}$ (41). Taking these concentrations, dividing by the duration of the experiments and extrapolating to the mass of the oceanic crust, we find that the corresponding global abiotic $CH_4$ estimates exceed the Earth's iron supply. As with McCollom 2013, these experiments suggest that high temperatures are necessary to generate $CH_4$. Such temperatures are higher than typical temperatures for habitable zone terrestrial planets. Therefore, we cannot properly extrapolate these experimental findings to a global abiotic $CH_4$ flux on temperate terrestrial planets.

**Impacts:**

- **Kasting 2005:** Kasting estimated the global $CH_4$ flux due to impact events during the Hadean to be 1.24 Tmol/year (35).

- **Kress & McKay 2004:** They determined that 0.6 Tmol of $CH_4$ is generated by a 1-km cometary impactor (42).

- **Zahnle et al. 2020:** For a highly-reduced Pluto-sized dwarf planet impactor, they determined that it would generate $\sim$2300 moles $CH_4$/cm$^2$ (11).

- **Court & Sephton 2009:** Experimentally studied ablation of carbonaceous chondritic materials and found that they release <100 ppm of $CH_4$ at temperatures up to 1000$^o$C (43).

**D. Calculations of Atmospheric Methane Lifetime for Volatile-Rich Bodies.** In Figure 5, we estimate the atmospheric lifetime of methane for an Earth-mass terrestrial planet with different water mass fractions and Titan-like initial volatile inventories. Using model calculations based on Cassini data for Titan's interior composition from (44), we assume Titan's volatile content consists of 0.35 % $CH_4$ and $\sim$4-6 % $CO_2$ relative to weight % $H_2O$ (44). We conservatively assume that the escape flux of hydrogen is diffusion-limited and calculate the atmospheric lifetime of $CH_4$. First we calculate the mass of methane:

$$m_{CH_4} = (0.35/100) m_{H_2O} \qquad [26]$$

where $m_{H_2O}$ is the water mass fraction of the planet (in kg). Then we calculate the diffusion-limited escape flux of $H_2$.

$$\Phi = C f_T(H_2) \qquad [27]$$

$\Phi$ is the escape flux of $H_2$ from Earth at the diffusion limit (in molecules/cm$^2$/s). Assuming the atmosphere is 10% $CH_4$, the fraction of hydrogen ($f_T(H_2)$) is 0.2 (i.e., $0.1 \times 2 = 0.2$ with two $H_2$ molecules per $CH_4$ molecule) and $C$ for Earth's atmosphere is $2.5 \times 10^{13}$ cm$^{-2}$s$^{-1}$. The atmospheric lifetime of $CH_4$ (in years) is given by:

$$T_{CH_4} = (\frac{m_{CH_4}}{M_{CH_4}} \text{NA}) / (\Phi \times \text{SA} \times 3.154 \times 10^7) \qquad [28]$$

where $M_{CH_4}$ is the molar mass of $CH_4$ in mol/kg, NA is the Avogadro constant and SA is the surface area of the Earth (in cm$^2$). Table S3 demonstrates how the lifetime of atmospheric $CH_4$ increases with increasing planetary mass fraction of water. For Figure 5, we ran a Monte Carlo simulation and varied the $CH_4$ inventory, sampling a uniform distribution from $10^{-4}$ to $10^{-2}$ relative to weight % water, for water mass fractions from $10^{-2}$ to 10 weight % of the planet's mass (assuming an Earth-mass planet).

To check that our estimated $CH_4$ atmospheric mixing ratio of 10% is reasonable, we calculate the solubilities of $CH_4$ and $CO_2$ for the atmosphere-ocean system reservoir using Henry's Law partitioning. For $CH_4$:

$$(C \times m_{H_2O} \times M_{CH_4}) + m_{\text{atm}} = m_{CH_4} \qquad [29]$$

where $m_{H_2O}$ and $m_{CH_4}$ are the masses of $H_2O$ and $CH_4$, respectively. $M_{CH_4}$ is $CH_4$'s molar mass (kg/mol) and $m_{\text{atm}}$ is the mass of the atmosphere, given by:

$$m_{\text{atm}} = PA/g \qquad [30]$$

where $P$ is pressure in bars, $A$ is the surface area of the planet in m$^2$, and $g$ is surface gravity (9.8 m/s$^2$). $[CH_4]$ is the concentration of dissolved $CH_4$ (mol/kg), which is given by Henry's Law:

$$[CH_4] = kP \qquad [31]$$

where $k$ is Henry's Law constant (0.0014 mol/kg/bar for $CH_4$). Solving for pressure, we find that for an Earth-mass planet with 1% of its mass consisting of water, the pressure of methane is ∼32 bars. Following the same formalism above for $CO_2$, which has a Henry's Law constant of 0.04 mol/kg/bar, its pressure is ∼22 bars. Therefore, our choice of $CH_4$'s atmospheric mixing ratio of 10% is conservative given the volatile inventories, which also allow for plausible inventories of $N_2$ gas.

We also consider whether volatile-rich, habitable zone planets could produce a long-lived $CH_4+CO_2$ biosignature false positive if not all water is melted. The storage and slow release of $CH_4$ and $CO_2$ from clathrates (ices that trap gases) on a Titan-like planet could conceivably mimic a biosphere. If surface conditions are habitable, however, then storage of large volumes of $CH_4$ in pure clathrates is not possible because $CH_4$ clathrates are less dense than liquid water at all pressures (45). Any $CH_4$ clathrates stored in high pressure ices would therefore rise to the surface and rapidly dissociate. It is true that methane clathrates are a $CH_4$ reservoir on Earth, but this is only because they are trapped by the weight of sediments above them, and are thus in a quasi-stable state (and will be potentially perturbed by slight surface warming). The weight of sediments could not trap the $\sim 10^{22}$ mol of $CH_4$ required to sustain biogenic-like fluxes of $CH_4$ for Gyr timescales.

If surface conditions are sub-freezing, $CH_4$ clathrates can inhibit subsurface ocean formation at all depths, and tectonically driven ice resurfacing may continuously bring fresh clathrates to the surface, maintaining $CH_4$ fluxes larger than Earth's biological flux (46). The region of parameter space for which atmospheric $CH_4$ can be maintained is likely small, however, since clathrates are unstable against surface warming: liquid water from warming will destabilize $CH_4$ clathrates, causing CH4 release into the atmosphere and even more greenhouse warming (46). Initial surface temperatures must therefore be low to prevent this runaway melting. For many planets, clathrate false positives may be ruled out by estimating minimum surface temperatures from observed atmospheric gases and plausible albedos. However, additional modeling work is required to characterize clathrate-atmosphere interactions across diverse planetary conditions.

| Input Parameter | Low | High | Sampling Method |
|---|---|---|---|
| Mantle Mass ($M_{\text{mantle}}$) (kg) | $0.1(4 \times 10^{24})$ | $10(4 \times 10^{24})$ | $log_{10}$ Uniform Distribution |
| Mantle $CO_2$ Mass ($M_{\text{mantle},CO_2}$) (kg) | $1 \times 10^{-5}(4 \times 10^{24})$ | $1 \times 10^{-2}(4 \times 10^{24})$ | $log_{10}$ Uniform Distribution |
| Mantle $H_2O$ Mass ($M_{\text{mantle},H_2O}$) (kg) | $1 \times 10^{-5}(4 \times 10^{24})$ | $1 \times 10^{-1}(4 \times 10^{24})$ | $log_{10}$ Uniform Distribution |
| Melt Fraction (F) | 0.1 | 0.5 | Uniform Distribution |
| Planetary Melt Production ($M_{\text{melt}}$) (g/s) | $0.1(3.2 \times 10^9)$ | $10(3.2 \times 10^9)$ | $log_{10}$ Uniform Distribution |

**Table S1. Monte Carlo sampling distributions for carbon partitioning and gas speciation calculations.**

Maggie A. Thompson, Joshua Krissansen-Totton, Nicholas Wogan, Myriam Telus, Jonathan J. Fortney

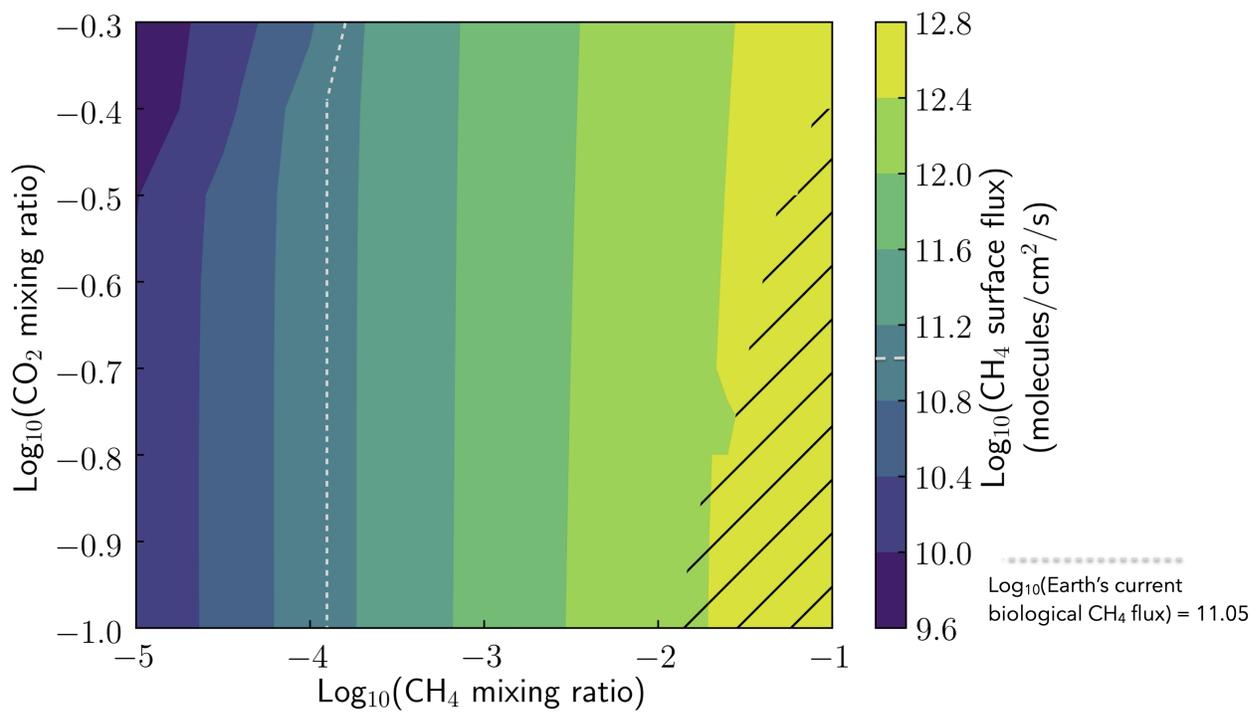

**Fig. S1. Methane surface flux required to sustain CH$_4$- and CO$_2$-rich atmospheres in photochemical steady state.** Using PhotochemPy, we ran a series of models with an initial atmospheric composition that is Archean Earth-like (orbiting the Sun at 2.7 Ga) exploring a range of CH$_4$ and CO$_2$ surface mixing ratios from $10^{-5}$ to 0.1 and 0.1 to 0.5, respectively. The contour colors correspond to the CH$_4$ surface flux required to sustain the atmospheric mixing ratios. While the model accounts for haze formation, we found that at higher CH$_4$ mixing ratios, the model had trouble converging to a steady-state solution. For those cases corresponding to the hatched region of the figure, we ran models that used the same Archean Earth-like initial atmospheric composition but removed the haze component in order to ensure model convergence. Ultimately, for abundant atmospheric CH$_4$ (i.e., surface mixing ratios above $\sim 10^{-3}$) to be stable against photochemistry in terrestrial planet atmospheres requires a significant replenishment source that results in large CH$_4$ surface fluxes that are likely much larger than Earth's current biological flux.

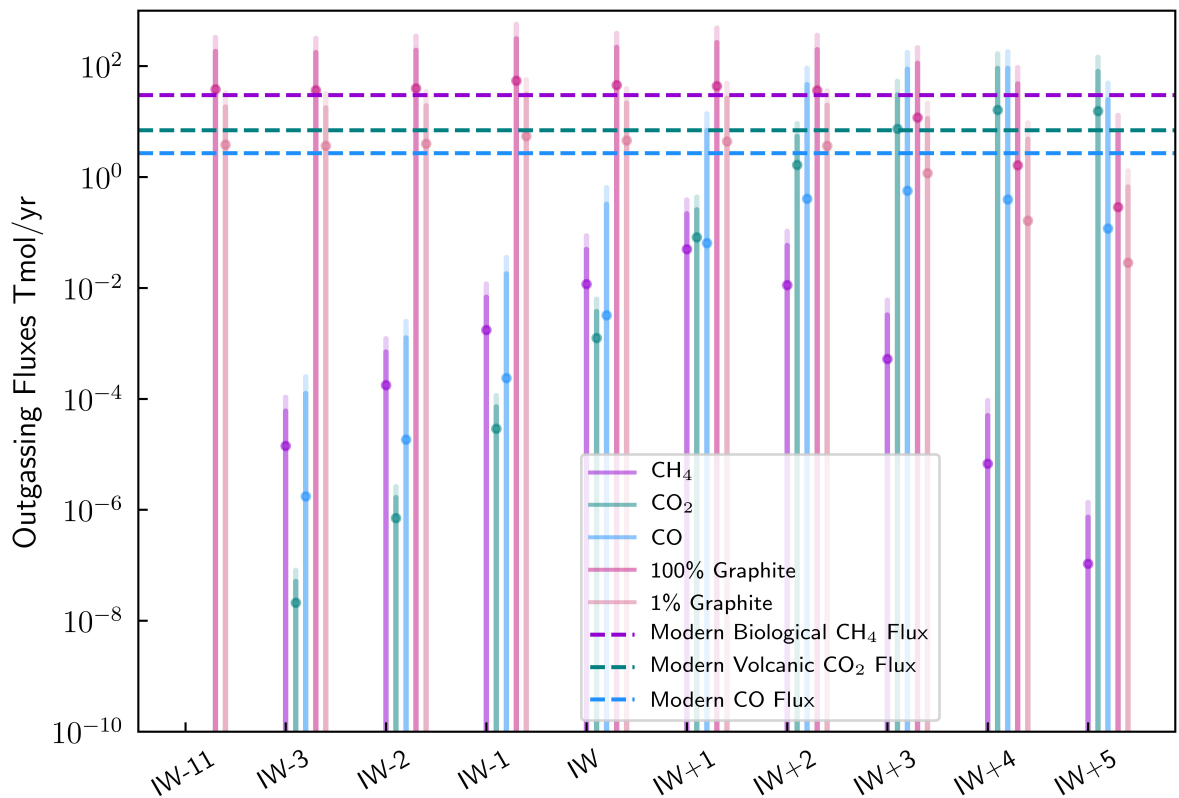

**Fig. S2. Simultaneous outgassing of CH$_4$ and CO$_2$ with negligible CO is highly unlikely unless large quantities of graphite are efficiently converted to CH$_4$ via metamorphism.** Outgassing fluxes as a function of oxygen fugacity. We used the same batch-melting model as described in Figure 3 and solved for speciation of gases produced by magmatic outgassing. The results are the average outgassing fluxes (in Tmol/year) of CH$_4$, CO$_2$ and CO from the Monte Carlo simulation with uncertainties reported as the 95% confidence intervals. The graphite results assume that either 100% or 1% of the remaining graphite can be converted into outgassed CH$_4$. The horizontal dashed lines show current outgassing fluxes on Earth for reference (e.g., biological CH$_4$ flux). For a planet with a very reduced melt composition, outgassing of any carbon species (i.e., CH$_4$, CO$_2$, and CO) will be negligible. In addition, for all oxygen fugacities considered from extremely reduced ($IW - 11$) to highly oxidized ($IW + 5$), the magmatic outgassing fluxes of CH$_4$ are still orders of magnitude lower than Earth's modern biological CH$_4$ flux of 30 Tmol/year.

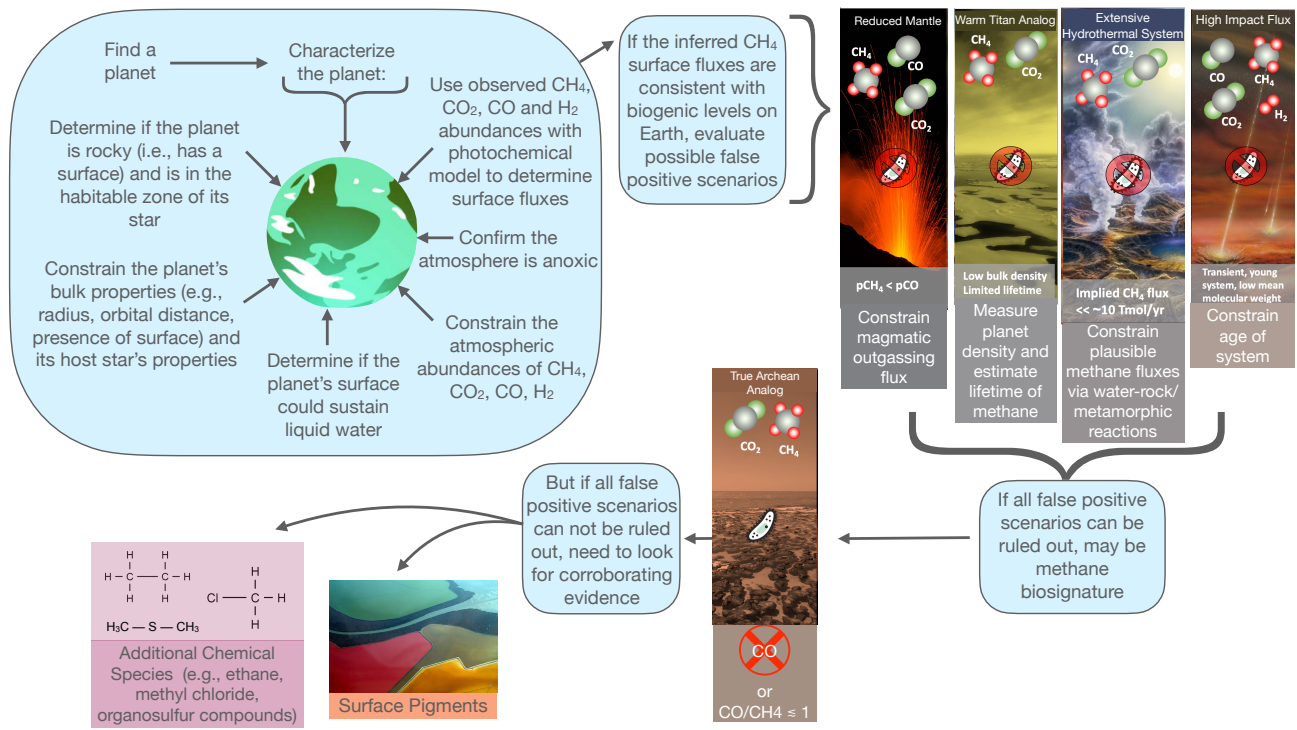

**Fig. S3. Possible procedure to search for methane biosignatures on terrestrial exoplanets that takes into account the planetary context.** Once an exoplanet has been detected, it is important to characterize its bulk properties (e.g., mass, radius, orbital properties, presence of a surface, host star properties). In addition, constraining its atmospheric composition, particularly the abundances $CH_4$, $CO_2$, CO, $H_2$, $H_2O$ and confirming that the atmosphere is anoxic, is essential for determining the presence of a methanogenic biosphere. Using this data with a photochemical model can determine the surface fluxes of the different atmospheric constituents that are necessary to sustain the observed atmospheric abundances. If the inferred $CH_4$ surface fluxes are consistent with plausible biogenic levels, then all possible false positive scenarios must be evaluated. If all false positives can be definitively ruled out then a methane biosignature has been identified at a high level of confidence that must be statistically determined. However, if all false positives cannot be ruled out, then it is necessary to look for corroborating evidence like additional gas species (e.g., methyl chloride, and organosulfur compounds) and the presence of surface pigments. Credits (images): Don Dixon, Wikimedia Commons; Donald Hobern; kuhnmi; NASA/JPL-Caltech/Lizbeth B. De La Torre; Doc Searls.

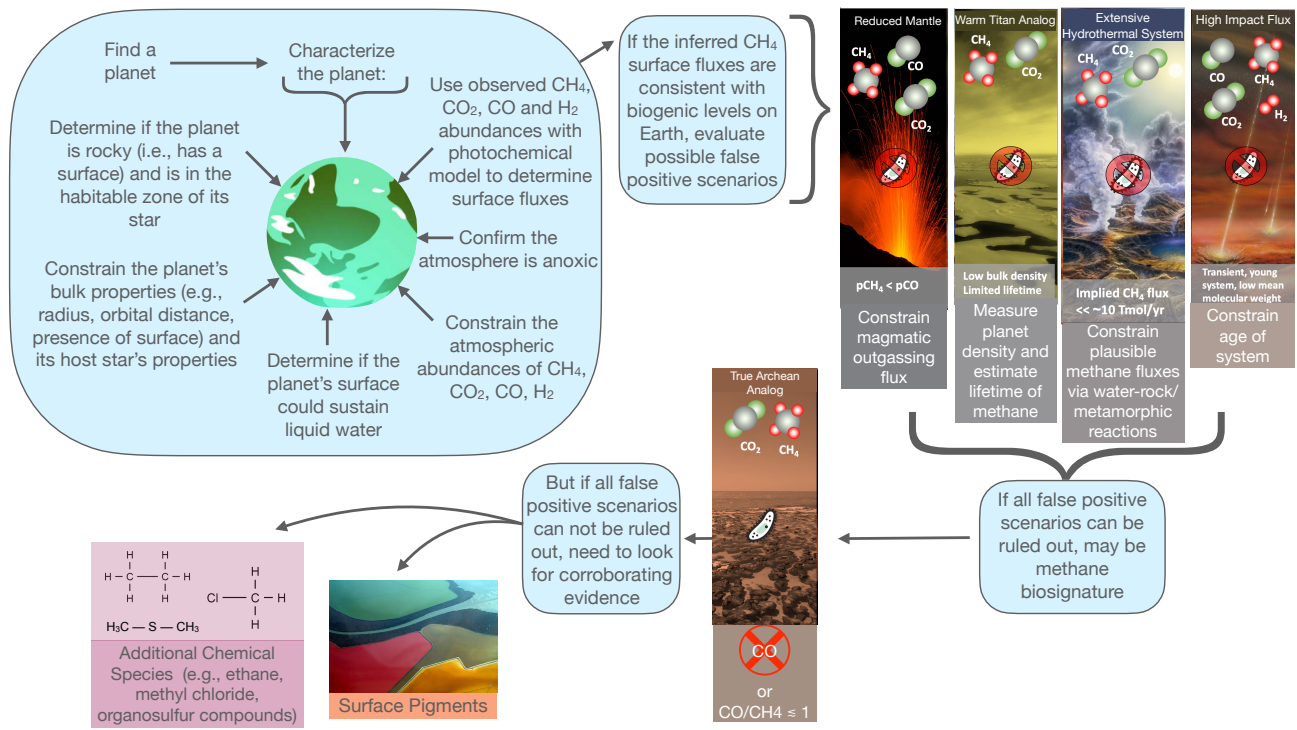

**Fig. S3. Possible procedure to search for methane biosignatures on terrestrial exoplanets that takes into account the planetary context.** Once an exoplanet has been detected, it is important to characterize its bulk properties (e.g., mass, radius, orbital properties, presence of a surface, host star properties). In addition, constraining its atmospheric composition, particularly the abundances $CH_4$, $CO_2$, CO, $H_2$, $H_2O$ and confirming that the atmosphere is anoxic, is essential for determining the presence of a methanogenic biosphere. Using this data with a photochemical model can determine the surface fluxes of the different atmospheric constituents that are necessary to sustain the observed atmospheric abundances. If the inferred $CH_4$ surface fluxes are consistent with plausible biogenic levels, then all possible false positive scenarios must be evaluated. If all false positives can be definitively ruled out then a methane biosignature has been identified at a high level of confidence that must be statistically determined. However, if all false positives cannot be ruled out, then it is necessary to look for corroborating evidence like additional gas species (e.g., methyl chloride, and organosulfur compounds) and the presence of surface pigments. Credits (images): Don Dixon, Wikimedia Commons; Donald Hobern; kuhnmi; NASA/JPL-Caltech/Lizbeth B. De La Torre; Doc Searls.

| Abiotic Source | Reference | CH$_4$ Flux | Global CH$_4$ Flux Estimate |
|---|---|---|---|
| **Higher Temperature** | | | |
| Volcanic | Etiope & Lollar 2013 | 100 tons/yr (Mt. Etna, most Icelandic volcanoes) | 9.4E-3 Tmol/year |
| Volcanic | Ryan et al. 2006 (29) | 9 tons/yr | 8.4E-4 Tmol/yr |
| Submarine Volcanism | (47) | — | 1.3E-2 Tmol/year |
| **Lower Temperature** | | | |
| Serpentinization at slow-spreading mid-ocean ridges | Cannat et al. 2010 (30) | 3.9E7 mol/yr (Rainbow Hydrothermal Field); 0.04-0.3E7 mol/km/yr (Ridge domains with frequent ultramafic outcrops) | 2.5E-2 Tmol/yr |
| Serpentinization in vent fluids from mid-ocean ridges | Keir 2010 (31) | 0.1-4 mmol/kg | 2E-2 Tmol/yr |
| Serpentinization at seafloor hydrothermal systems | Catling & Kasting 2017 (4) | — | 0.015 - 0.03 Tmol/year |
| Serpentinization at hydrothermal vent systems | Guzmán-Marmolejo et al. 2013 (34) | — | 0.18 Tmol/year (for 1 M$_{Earth}$), 0.35 Tmol/year (for 5 M$_{Earth}$) |
| Serpentinization at off-axis mid-ocean ridges | Kasting 2005 (35) | — | 0.1 Tmol/year (at present), 1.5 Tmol/year (during Hadean) |
| Serpentinization and carbonate reduction at subduction zones | Brovarone et al. 2017 (36) | — | 9E-2 Tmol/yr (subduction zone fluids); 8E-3-0.2 Tmol/yr (forearc mantle wedges above subduction zones); 1E-2-0.1 Tmol/yr (sub-seafloor) |
| Serpentinization at subduction-related sites (and estimates for Archean Eon) | Fiebig et al. 2007 (37) | — | 2.5-5 Tmol/yr (during Archean), 6E-3 Tmol/yr (at present) |
| Reduction of CO$_2$ in continental hydrothermal systems | Fiebig et al. 2009 (38) | — | 0.31 Tmol/year |
| Serpentinization of chromitites in ophiolites | Portella et al. 2019 (39) | 0.31 $\mu$g/g (rock) in chromitites | 2E-3 Tmol/yr |
| Serpentinization in subduction zones, mid-ocean ridges and ophiolites | Klein et al. 2019 (40) | 2E6-11E6 mol/yr (Chimaera system) | 1.5E-6 Tmol/yr |
| Experiments of hydrocarbon formation in deep subsurface | McCollom 2013 (33) | 0.05 $\mu$mol/kg/hr | CH$_4$ Flux > Fe supply |
| Serpentinization experiments for mid-ocean ridges and forearcs | Jones et al. 2010 (32) | 1E-5 - 0.06 $\mu$mol/kg/hr | CH$_4$ Flux > Fe supply* |
| Serpentinization experiments investigating mineral catalysts | Oze et al. 2012 (48) | 0.15 $\mu$mol/kg/hr | CH$_4$ Flux > Fe supply* |
| Serpentinization experiments on forsteritic olivine | Neubeck et al. 2011 (49) | 2.7E-11-7.3E-11 mol/m$^2$/s | 0.64-1.2 Tmol/yr* |
| Serpentinization experiments on olivine | McCollom 2016 (41) | ~7.7E-3-1.3E-2 $\mu$mol/kg$_{olivine}$/hr | CH$_4$ Flux > Fe supply |
| **Impacts** | | | |
| Impact events during the Hadean | Kasting 2005 (35) | — | 1.24 Tmol/year |
| Cometary impact events | Kress & McKay 2004 (42) | 0.6 Tmol (generated by 1-km comet impactor) | — |
| Impact events for early Earth | Zahnle et al. 2020 (11) | 2300 moles/cm$^2$ (generated by a highly-reduced Pluto-sized dwarf planet impactor) | Not applicable - transient event lasting ~10,000 years |
| Meteorite ablation experiments | Court & Sephton 2009 (43) | <100 ppm from gasification of carbonaceous asteroid | — |

Table S2. Summary of abiotic CH$_4$ sources and their estimated global CH$_4$ flux values. *Indicates that the experimental measurements may have over-estimated the amount of methane generated due to the presence of background sources.

Maggie A. Thompson, Joshua Krissansen-Totton, Nicholas Wogan, Myriam Telus, Jonathan J. Fortney

| Mass Fraction of Water (wt% of planet mass) | 0.1 | 1.0 | 10 | 50 |
|---|---|---|---|---|
| Lifetime of $CH_4$ (Myr) | 1 | 10 | 100 | 500 |

**Table S3. Estimated lifetime of atmospheric $CH_4$ for Earth-mass terrestrial planets with Titan-like initial volatile inventory and different size water mass fractions.**